\documentclass{pasj01}

\Received{}
\Accepted{}
\Published{}

\usepackage{xcolor}
\usepackage{booktabs}
\usepackage{graphicx}

\begin{document}

\title{Star Cluster Formation in Orion A}

\author{{Wanggi Lim}\altaffilmark{1},
{Fumitaka Nakamura}\altaffilmark{2},
{Benjamin Wu}\altaffilmark{2},
{Thomas G. Bisbas}\altaffilmark{3,4,5},
{Jonathan C. Tan}\altaffilmark{6,7},
{Edward Chambers}\altaffilmark{1},
{John Bally}\altaffilmark{8},
{Shuo Kong}\altaffilmark{9},
{Peregrine McGehee}\altaffilmark{10},
{Dariusz C. Lis}\altaffilmark{11},
{Volker Ossenkopf-Okada}\altaffilmark{4},
{\'{A}lvaro S\'{a}nchez-Monge}\altaffilmark{4}
}
\altaffiltext{1}{SOFIA-USRA, NASA Ames Research Center, MS 232-12, Moffett Field, CA 94035, USA}
\altaffiltext{2}{National Astronomical Observatory of Japan, Mitaka, Tokyo 181-8588, Japan}
\altaffiltext{3}{National Observatory of Athens, Institute for Astronomy, Astrophysics, Space Applications and Remote Sensing, Penteli, 15236, Athens, Greece}
\altaffiltext{4}{I. Physikalisches Institut, Universit\"at zu K\"oln, Z\"ulpicher Stra{\ss}e 77, Germany}
\altaffiltext{5}{Department of Physics, Aristotle University of Thessaloniki, GR-54124 Thessaloniki, Greece}
\altaffiltext{6}{Department of Space, Earth \& Environment, Chalmers University of Technology, Gothenburg, Sweden}
\altaffiltext{7}{Department of Astronomy, University of Virginia, Charlottesville, VA 22904, USA}
\altaffiltext{8}{Department of Astrophysical and Planetary Sciences, University of Colorado, Boulder, CO, USA}
\altaffiltext{9}{Steward Observatory, University of Arizona, AZ 85719}
\altaffiltext{10}{Department of Earth and Space Sciences, College of the Canyons, Santa Clarita, CA 91355, USA}
\altaffiltext{11}{Jet Propulsion Laboratory, California Institute of Technology, 4800 Oak Grove Drive, Pasadena, CA 91109, USA}

\email{wlim@usra.edu}


\maketitle

\begin{abstract}
We introduce new analysis methods for studying the star cluster formation processes in Orion~A, especially examining the scenario of a cloud-cloud collision. We utilize the CARMA-NRO Orion survey $^{13}$CO~(1-0) data to compare molecular gas to the properties of YSOs from the SDSS~III IN-SYNC survey. We show that the increase of $v_{\rm 13CO} - v_{\rm YSO}$ and $\Sigma$ scatter of older YSOs can be signals of cloud-cloud collision. SOFIA-upGREAT 158$\micron$ [C{\sc ii}] archival data toward the northern part of Orion~A are also compared to the $^{13}$CO data to test whether the position and velocity offsets between the emission from these two transitions resemble those predicted by a cloud-cloud collision model. We find that the northern part of Orion~A, including regions ONC-OMC-1, OMC-2, OMC-3 and OMC-4, shows qualitative agreements with the cloud-cloud collision scenario, while in one of the southern regions, NGC1999, there is no indication of such a process in causing the birth of new stars. On the other hand, another southern cluster, L1641N, shows slight tendencies of cloud-cloud collision. Overall, our results support the cloud-cloud collision process as being an important mechanism for star cluster formation in Orion~A.

\end{abstract}

\section{Introduction}

Star cluster formation is an essential mechanism for the evolution of galaxies. Especially energetic feedback from massive stars associated with young clusters is likely to play a dominant role in determining the morphology and star formation efficiency of the natal giant molecular clouds (GMCs). Despite this importance, we have quite limited theoretical understanding and observational constraints of how massive stars and star clusters form, since these systems are typically found far from the Sun, i.e., several kpcs, and are deeply embedded inside dense molecular structures at their early evolutionary stages.

It is widely accepted that most stars form in clusters \citep{2003ARA&A..41...57L, 2009ApJS..184...18G} from the densest parts of GMCs (so called molecular clumps, \cite{2000prpl.conf...97W}). \citet{2007prpl.conf..165B} proposed three simple evolutionary stages of massive star cluster formation -- massive starless clumps, protoclusters and final star clusters. Infrared dark clouds (IRDCs, \cite{1996A&A...315L.165P,1998ApJ...508..721C}) are promising sources in which to search for such massive early stage clumps and cores (e.g., \cite{2005ApJ...630L.181R,2009ApJS..181..360C,2012ApJ...754....5B,tan14,2016ApJ...821...94K,2018ApJS..236...25K,2018ApJ...862..105L,2018ARA&A..56...41M}). On the other hand, Giant HII regions are representative sources to investigate the protoclusters harboring already-formed massive stars \citep{2002ARA&A..40...27C,2004MNRAS.355..899C,2007prpl.conf..181H,2019ApJ...873...51L,2019A&A...624A..63B}. However, even in the recent studies of IRDCs and Giant HII regions, observational limitations of angular resolution affect the ability to determine detailed interstellar medium (ISM) structures (e.g., \cite{2015A&A...578A..29S,2015A&A...573A.119R}) and to distinguish individual YSOs in young clusters \citep{2019ApJ...873...51L}.

\begin{figure*}
 \begin{center}
  \includegraphics[width=6.0in]{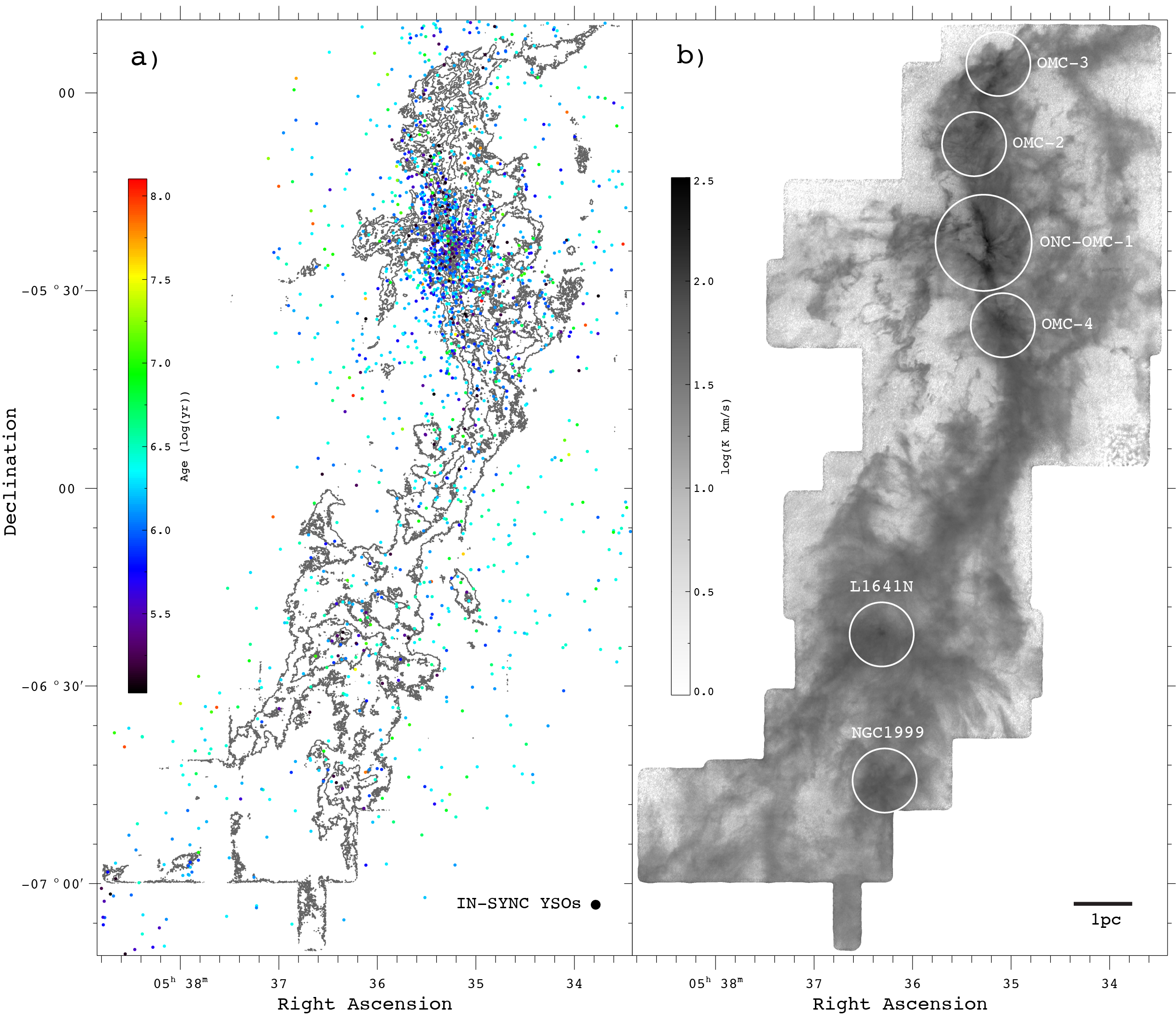}
 \end{center}
 \caption{{\bf Left:} The 0th moment map of CARMA-NRO $^{13}$CO(1-0) (contours) toward Orion A, with SDSS-IN-SYNC YSO positions marked with dots. The colors indicate YSO ages. {\bf Right:} The $^{13}$CO~(1-0) 0th moment map with star-forming clumps studied by \citet{2018ApJS..236...25K} shown by white circles. 
 }\label{fig:ovly}
\end{figure*}

\begin{figure}
 \begin{center}
  \includegraphics[width=2.5in]{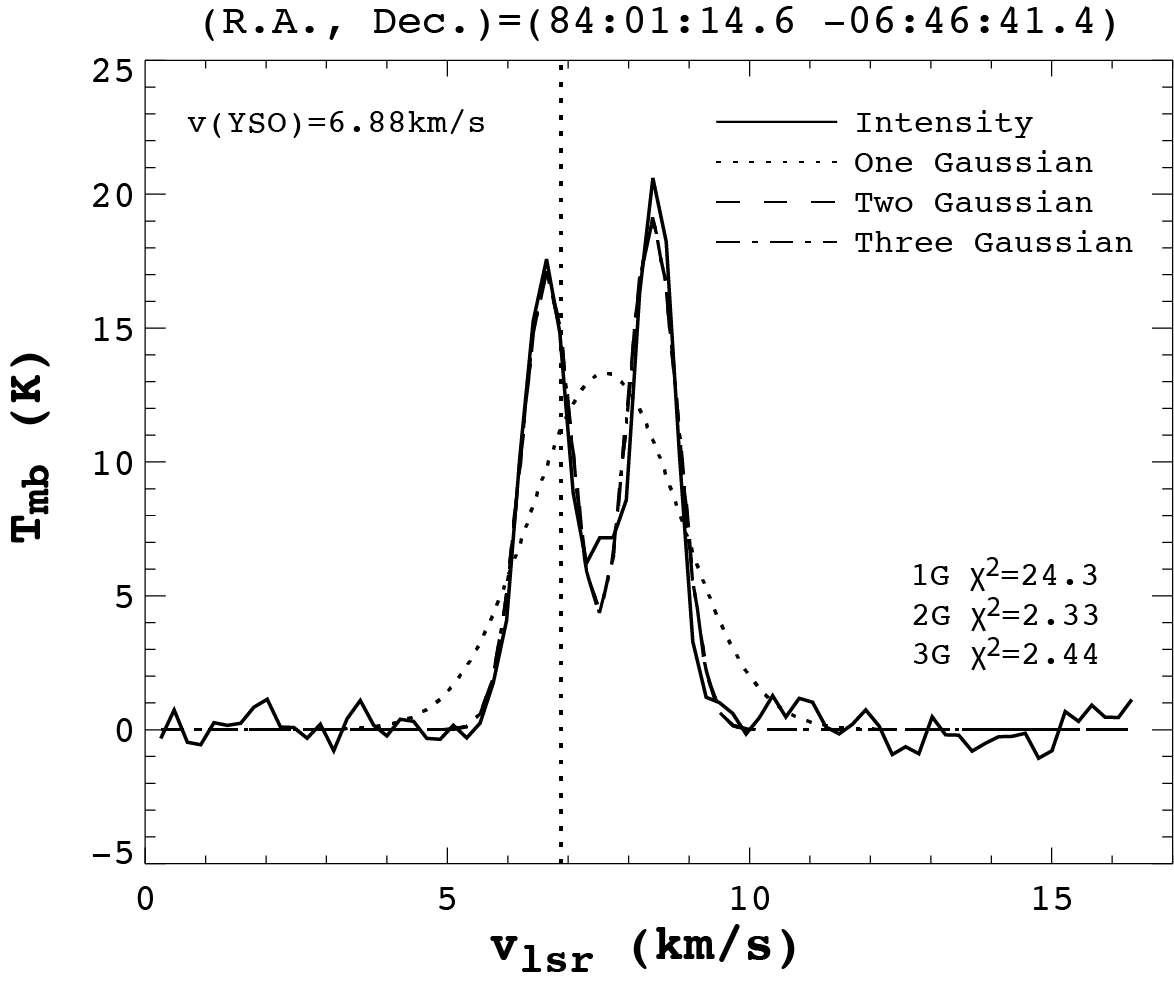}
 \end{center}
 \caption{
An example $^{13}$CO(1-0) spectrum at an IN-SYNC \citep{2016ApJ...818...59D} YSO position with multi-Gaussian fits (R.A. and Dec. at the top of the plot). Solid lines show the observed $^{13}$CO(1-0) spectrum. The black dotted, dashed and long-dashed lines show single, double and triple Gaussian fits to the spectrum. We select the fiducial multi-Gaussian model from the minimum $\chi^2$ values among the models (in this case, two Gaussian fit). The vertical dotted line indicates the YSO velocity. We derive the $\sigma$ values of all 1,371 spectra (sources in the CARMA-NRO mapping area) and adopt the median value, $\sigma\sim0.59$~K, to determine the intensity peaks at $T_{\rm mb}\gtrsim3\sigma\sim1.8$~K.}\label{fig:spec}
\end{figure}

Orion~A is the closest massive star-forming cloud at a distance of $\sim$400~pc from the Sun \citep{2007A&A...474..515M,2017ApJ...834..142K, 2018AJ....156...84K},  which is about an order of magnitude closer than typical IRDCs and Giant HII regions \citep{2004MNRAS.355..899C,2006ApJ...653.1325S,2013ApJ...770...39E}. Orion~A is also recognized for its similarities to a well-known IRDC, G11.11-0.12 \citep{2013A&A...557A.120K,2015A&A...573A.119R} and the presence of active HII regions \citep{2016ApJ...830..118S}, making it the best laboratory to investigate cluster formation containing a wide range of evolutionary stages and with stars spanning a wide range of masses. It is essential to study the kinematics of such structures to test dynamical models of the star formation process. 

Two examples of the recent studies of molecular structures and young stellar objects (YSOs) in Orion~A are the CARMA-NRO Orion Survey \citep{2018ApJS..236...25K} and the INfrared Spectra of Young Nebulous Clusters (IN-SYNC) project \citep{2016ApJ...818...59D}. \citet{2018ApJS..236...25K} achieved the highest angular resolution $^{12}$CO(1-0), $^{13}$CO(1-0) and C$^{18}$O(1-0) observations to date for Orion~A ($\sim$5$\arcsec$) by merging Combined Array for Research in Millimeter Astronomy (CARMA) and Nobeyama Radio Observatory 45 m telescope (NRO) data.
One needs to note that the best CO observations ($J$=1-0) toward Orion~A before the CARMA-NRO Orion Survey were from the NRO survey.
Using the NRO only CO data with a moderate angular resolution of $\sim 20\arcsec$, \citet{2018ApJ...862..121F}, \citet{2019PASJ..tmp...87I}, and \citet{tanabe19} analyzed molecular cloud structure and particularly investigated effects of stellar feedback in Orion A.
As a part of the IN-SYNC project, \citet{2016ApJ...818...59D} examined kinematic associations between molecular structures and YSOs in Orion~A by comparing $^{13}$CO(1-0) position-velocity diagrams (PVDs) with the radial velocities of $\sim$2,700 YSOs.

GMC-GMC collisions are potential triggers of star formation (e.g., \cite{2000ApJ...536..173T}). In fact, star cluster formation in Orion~A has been suggested to be the result of GMC collisions by many observational studies (e.g., \cite{2012ApJ...746...25N,2018ApJ...859..166F}). However, these studies have typically been limited to searching for the "bridge effect" \citep{2015MNRAS.450...10H, 2017ApJ...850...23B} in position-velocity diagrams from molecular gas tracers, such as $^{12}$CO(1-0). Recently, \citet{2018MNRAS.478L..54B} introduced a new method to test the observational evidence of GMC collisions by examining both $^{13}$CO(1-0) emission (from the Galactic Ring Survey [GRS]; \cite{2006ApJS..163..145J}) and 158$\micron$ [C{\sc ii}] emission (from SOFIA-upGREAT) toward an IRDC. They found position and velocity offsets between $^{13}$CO(1-0) and [C{\sc ii}] intensity peaks of IRDC G035.39-00.33, to be consistent with those derived in simulations and synthetic observations of cloud-cloud collisions. 

In this study, we investigate star cluster formation in Orion~A by utilizing the $^{13}$CO CARMA-NRO Orion Survey data \citep{2018ApJS..236...25K} and IN-SYNC YSO information \citep{2016ApJ...818...59D}. For a more detailed look at potential cloud-cloud collision signatures in the Orion~A region, we also utilize a recent SOFIA-upGREAT [C{\sc ii}] observation \citep{2019Natur.565..618P} originally designed to study the effect of stellar feedback. We compare the observational analyses to theoretical models. 

The paper is structured as follows. In section~\ref{sec:method1}, we address the analysis methods of the observational data. Section~\ref{sec:results} presents the results from this analysis. 
We discuss the observational results and their comparison to theoretical models in section~\ref{sec:discussions}.

\section{Observational Data Analysis Methods}\label{sec:method1}

\subsection{Summary of the SDSS IN-SYNC Data}

The IN-SYNC survey was performed with the SDSS-III Apache Point Observatory Galactic Evolution Experiment (APOGEE), a fiber-fed multi object infrared spectrograph, operating in $H$-band in the wavelength range $1.5\micron\lesssim\lambda\lesssim1.7\micron$ with $2\arcsec$ size fibers. The spectral resolution is $\lambda/\Delta\lambda\simeq$22,500 and the observation had a selection limit at 12.5~$H$~magnitude.
\citet{2016ApJ...818...59D} analysed $\sim$2,700 sources in the area of the Orion~A cloud and identified $\sim$1,900 YSOs as members of the Orion~A cloud. 

The IN-SYNC data provide estimates of stellar parameters and radial velocities of YSOs in the Orion~A area. In our study we focus on the estimates of age and radial velocity of the YSOs. The median uncertainties of the age are $\sim$1.15~Myr and for radial velocity are $\sim$0.91~km/s. The left panel of figure~\ref{fig:ovly} shows the location of IN-SYNC YSOs with the colors related to their different age.

\subsection{Summary of the CARMA-NRO Orion Survey and Methods of Analysis for $^{13}$CO(1-0)}

The CARMA-NRO Orion Survey provides combined $^{13}$CO(1-0) data with Full Width Half Maximum (FWHM) of the beam $\simeq8\arcsec\times6\arcsec$, position angle $\simeq10\degree$, channel width $\simeq0.22$~km~s$^{-1}$, and rms noise level $\simeq$ 0.64~K (\cite{2018ApJS..236...25K}; see also \cite{2019PASJ..tmp...32N} for the details of the Nobeyama observations). The map spans approximately 2.3$\degree$ along the declination direction covering from OMC-3 (northernmost cloud) to NGC1999 (southernmost cloud) as shown in the right panel of figure~\ref{fig:ovly}.

Here, we take advantage of the high angular resolution of these $^{13}$CO(1-0) observations, which is generally sufficient to give independent measurements of gas properties around each YSO in the IN-SYNC survey. We compare the $^{13}$CO(1-0) spectra from individual pixels containing IN-SYNC YSO positions to the stellar properties of the corresponding YSOs. There are 1,529 YSOs defined as members of Orion~A filamentary structure along the Declination range of the CARMA-NRO Orion Survey mapping area. Among the IN-SYNC YSOs, 1,371 sources are located where the $^{13}$CO(1-0) data exist. We investigate all of 1,371 $^{13}$CO(1-0) spectra and compare them with the YSO properties. We will focus on exploring the physical and kinematic relation between YSOs and the molecular gas (hereafter Gas-YSO association). 

Most of the investigated $^{13}$CO(1-0) spectra ($\sim87\%$) show multiple velocity components (typically 2--3). We have developed a simple automated 
metric to determine the number of components and their properties through a multiple Gaussian component fitting procedure. For the example spectrum shown in figure~\ref{fig:spec}, we plot the best-fit reduced $\chi^2$ so that $\chi_{\nu}^2 = [\sum\chi^2]/{N_{\rm DOF}}$ where $\chi^2$ = [($T_{\rm mb,obs}-T_{\rm mb,exp}$)/$\sigma$]$^2$ for each channel and {${N_{\rm DOF}}$} (the number of degrees of freedom) $= N_{\rm channel}-N_{\rm par}$. $T_{\rm mb,obs}$ is the observed $^{13}$CO(1-0) brightness temperature of the channel, $T_{\rm mb,exp}$ is the expected $^{13}$CO brightness temperature of the channel, the total number of channels ($N_{\rm channel}$) is 74, and the number of parameters ($N_{\rm par}$) are 3, 6, or 9 (for single, double or triple Gaussian cases, respectively). We determine the 1~$\sigma$ noise level of the $^{13}$CO spectra by examining line-free velocity ranges (typically $0-5$km/s) of all 1,529 sources and derive the median value which is $T_{\rm mb}\sim$0.59~K. We then select the line emission components over 3~$\sigma$ ($T_{\rm mb}\sim1.8$~K). 
The $\chi_{\nu}^2$ values are compared with single, double and triple Gaussian models, then the best model is selected that minimizes $\chi_{\nu}^2$. 

We compare our $^{13}$CO(1-0) fitting method to that of the Semi-automated multi-COmponent Universal Spectral-line fitting Engine (SCOUSE, \cite{2016MNRAS.457.2675H}). SCOUSE determines, with user input help, the number of velocity components and their profile information for a given spectrum. We compare peak intensities, widths, and centroids, between both methods for 50 selected $^{13}$CO spectra. For all of the 50 spectra, our fitting method agrees well with SCOUSE. We note that our method and SCOUSE both utilize MPFIT (a non-linear least squares fitting methods; \cite{2009ASPC..411..251M}) with same concepts of $N_{\rm DOF}$ and $\chi_\nu^2$. Thus, it is not surprising to see good agreement between the two methods. We also compare the multi-component fitting models of this study to a PYTHON based fully automated fitter for multi-component line emissions, the Behind The Spectrum (BTS; \cite{2018MNRAS.479.1722C}). The difference between SCOUSE and BTS are the input of user-defined parameters where BTS only needs minimal inputs. We show the detail comparison of the derived parameters of the 50 sample spectra with the three different methods in Appendix~\ref{sec:app}. The comparison shows that the investigation of the main $^{13}$CO(1-0) velocity components of the fitting methods of this study is very robust.

The properties of individual YSOs (age and radial velocity, $v_{\rm YSO}$; \cite{2016ApJ...818...59D}) are compared to the major $^{13}$CO(1-0) velocity structures (containing the strongest peak) of the corresponding locations (i.e., $v_{\rm 13CO}$): in particular, we examine the difference in radial velocity between the molecular structures as traced by $^{13}$CO(1-0) and the YSOs ($\Delta v=v_{\rm 13CO}-v_{\rm YSO}$) at each YSO position. Then, we search for any trend in $\Delta v$ with YSO age.

\begin{table*}
\tbl{Derived Parameters of Gas-YSO Association for entire Orion~A}{
\begin{tabular}{l|cccccccc}
\hline\noalign{\vskip3pt} 
                        & Age bin (yrs)& $<10^6$ & $10^6$ to $10^{6.5}$ & $10^{6.5}$ to $10^{7.0}$ & $10^{7.0}$ to $10^{7.5}$ & $>10^{7.5}$ \\
\hline\noalign{\vskip3pt}
Detected        & $\Delta v$ (km/s)         &0.28&0.78&-0.16&-0.12&-0.52\\
                & $\sigma (\Delta v$) (km/s)&5.48&4.87& 9.59&18.20&14.92\\
                & Normalized $\Sigma$       &0.05&0.03& 0.03& 0.03& 0.02\\
\hline\noalign{\vskip3pt} 
Non detection    & $\Delta v$ (km/s)         &3.39&0.67& 2.19& 3.63& 7.35\\
                & $\sigma (\Delta v$) (km/s)&8.60&6.16& 3.33&11.95& 9.67\\
\hline\noalign{\vskip3pt} 
Total           & $\Delta v$ (km/s)         &0.35&0.76&-0.10& 0.43&-0.52\\
                & $\sigma (\Delta v$) (km/s)&5.61&5.05& 9.13&16.96&14.09\\
\hline\noalign{\vskip3pt} 
\end{tabular}}\label{tb:prmgl}
\begin{tabnote}
\end{tabnote}
\end{table*}

We also measure the mass surface density of the molecular gas ($\Sigma$) following the method of, e.g., \citet{2011ApJ...730...44H}, i.e., starting from the basic theoretical assumptions:
\begin{equation}\label{eqn:n13}
\frac{dN_{\rm 13CO}}{d\nu}=\frac{8\pi}{A\lambda_{\rm 0}^3}\frac{g_l}{g_u}\frac{\tau_\nu Q_{\rm rot}}{1-{\rm exp}(-h\nu/kT_{\rm ex})},
\end{equation}
\noindent where $\lambda_0=$0.272~cm for $^{13}$CO(1-0), $g_l$ and $g_u$ are the statistical weights of the lower and upper levels, $A$ is the Einstein A coefficient (6.3355 $\times$ 10$^{-8}$ s$^{-1}$), $T_{\rm ex}$ is the excitation temperature, $h$ is the Planck constant, and $k$ is the Boltzmann constant. $Q_{\rm rot}$ is the partition function which for linear molecules is given by $Q_{\rm rot}=\Sigma^\infty_{J=0}(2J+1){\rm exp}(-E_J/kT_{\rm ex})$ with $E_J=J(J+1)hB$, where $J$ is the rotational quantum number and $B$ is the rotational constant for $^{13}$CO(1-0). For the $T_{\rm ex}$ of each pixel, we utilize the $^{12}$CO $T_{\rm ex}$ map derived by \citet{2018ApJS..236...25K} as the $^{13}$CO $T_{\rm ex}$. 
The $\tau_\nu$ is the line optical depth, which is integrated over corresponding velocity/frequency ranges, and is estimated from the relationship
\begin{equation}\label{eq:tau}
T_{B,\nu}=\frac{h\nu}{ k}[{ f}({ T}_{\rm ex})-{ f}({ T}_{\rm bg})][1-{ e}^{-\tau_{\nu}}], 
\end{equation}
\noindent where $T_{B,\nu}$ is the main beam brightness temperature at frequency $\nu$, $f(T)$=[exp($h\nu/[kT])-1]^{-1}$ and $T_{\rm bg}$ is the background temperature of 2.725~K. The CO isotopic column densities are converted to total mass surface densities using the $^{13}$CO/$^{12}$CO and $^{12}$CO/H$_2$ abundance ratios from \citet{2005ApJ...634.1126M} and \citet{1994ApJ...428L..69L}, respectively. \citet{2005ApJ...634.1126M} assumed the $^{13}$CO/$^{12}$CO ratio to be ${{X}_{\rm 12CO}}/{{X}_{\rm 13CO}}=6.2\times{{D}_{\rm GC}}/{\rm kpc}+18.7$, where D$_{\rm GC}$ is Galactocentric distance. The Orion~A region has ${{n}_{\rm 12CO}}/{{n}_{\rm 13CO}}\simeq$69.6 (c.f. 62$\pm$4; \cite{1993ApJ...408..539L}, and 76$\pm$6; \cite{2015ApJ...813..120S}). \citet{1994ApJ...428L..69L} estimated $^{12}$CO abundance, ${{X}_{\rm 12CO}}=2.0 \times 10^{-4}$.
Based on these abundance relations, we calculate the mass surface density of each pixel as, $\Sigma={\mu_{\rm H}}\times N_{\rm 13CO}/{{X}_{\rm 12CO}}$\,${\rm g}\ {\rm cm^{-2}}$, where $\mu_{\rm H}$=2.34$\times$10$^{-24}$~g is the mean molecular mass in the gas. One needs to note that $N_{\rm 13CO}$ is derived by integrating over the major velocity components only. The Gas-YSO association is investigated by comparing the $\Sigma$ of the $\sim$1,500 YSO positions to the ages of corresponding YSOs.

\subsection{Summary of SOFIA-upGREAT [C{\sc ii}] Data}

We utilize a publicly available SOFIA-upGREAT 158$\micron$ [C{\sc ii}] data set that stems from observations of the northern part of Orion~A cloud \citep{2019Natur.565..618P}. The 158~$\micron$ SOFIA-upGREAT data have an angular resolution of $\sim18\arcsec$. The [C{\sc ii}] data cube has a velocity resolution of $\sim0.2$~km/s, and an rms noise level of $\sim$1.0~K ($T_{\rm mb}$). 

In this work, we perform a qualitative analysis to understand the dynamics of the Orion~A region, i.e., whether or not a cloud-cloud collision scenario can be supported. We, thus, perform normalizations to explore the \emph{trends} as indicated in figure~\ref{fig:pltgb}.

\section{Results}\label{sec:results}

\subsection{Correlations between gas and YSOs}\label{sec:result1}

\subsubsection{'$\Delta v$~vs.~age' and '$\Sigma$~vs.~age' of entire Orion~A cloud}

Figure~\ref{fig:ovly}(a) shows the location of YSOs on top of the Orion~A molecular gas structures. The color of the YSO represents its age \citep{2016ApJ...818...59D}, increasing from blue (younger) to red (older). It should be pointed out that YSO ages based on stellar isochrones are quite uncertain, perhaps by a factor of up to 2 (see, for example, the discussion in \cite{2011ApJ...732....8V}, \cite{2014prpl.conf..219S} and \cite{2016ApJ...818...59D}). While there is likely to be a minimum uncertainty level of at least 1~Myr in any given absolute age estimate, relative ages are likely to be better determined.

\begin{figure}
 \begin{center}
\includegraphics[width=2.5in]{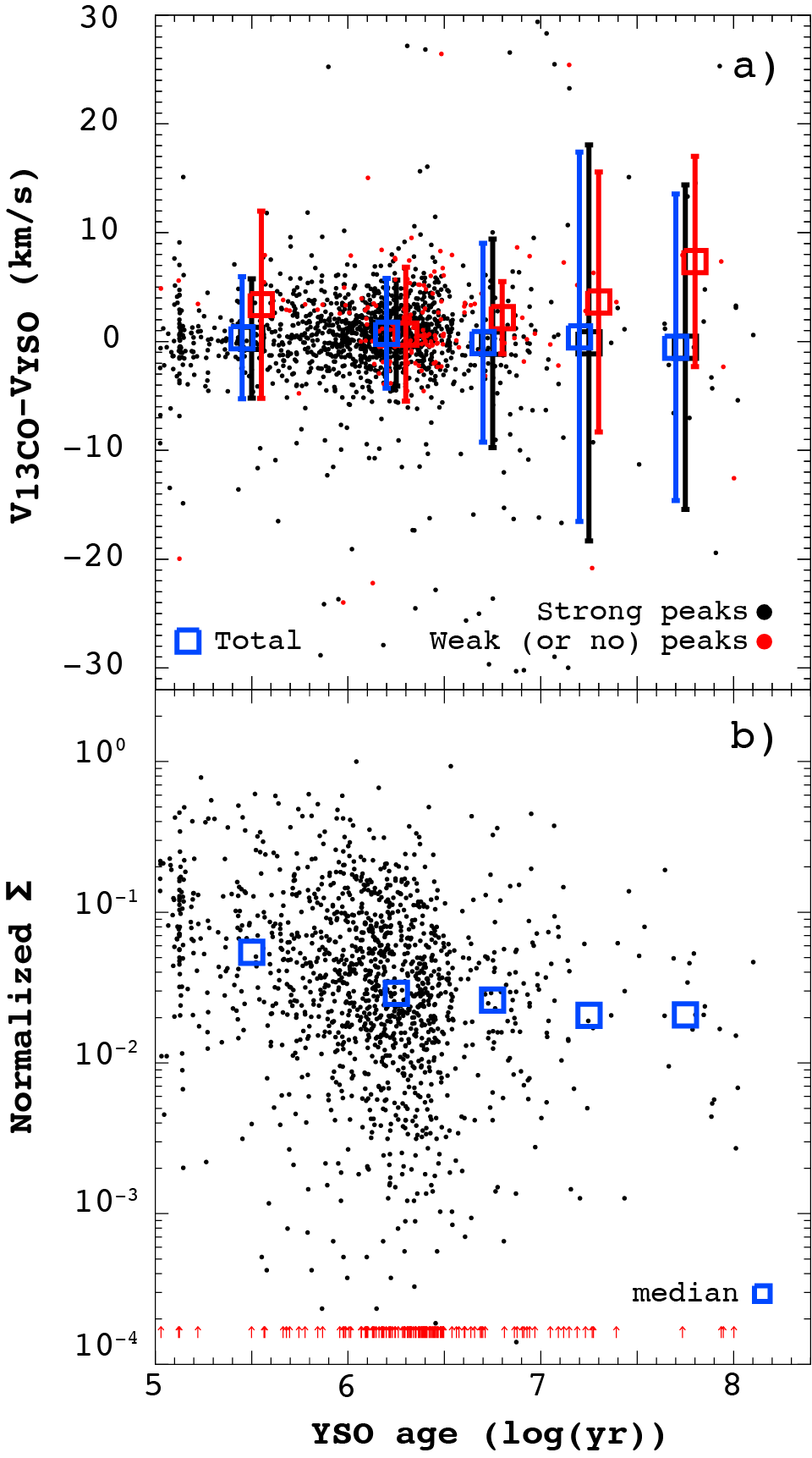}
 \end{center}
 \caption{{\bf Top:} $\Delta v$ vs.~YSO age of Orion~A. The black dots are the YSOs where $^{13}$CO(1-0) profiles are detected at the pixel position. The red dots are the YSOs with weakly or not detected $^{13}$CO(1-0) lines. Average $^{13}$CO(1-0) line profile around the pixel position is used for each red dotted YSO. The black, blue and red boxes indicate the median velocity differences in each YSO age bin (see text for details) for black, red and all dots, respectively. The velocity differences are derived from the $^{13}$CO(1-0) velocity component containing the major peak. {\bf Bottom:} $\Sigma$ vs. YSO~age of Orion\,A. The black dots are same as the Top panel. We normalize the $\Sigma$ values so that the maximum normalized $\Sigma$ among the all YSOs in Orion~A area is 1. Red upward arrow present the YSO ages of the red dotted data points in the Top panel. Blue boxes show the median normalized $\Sigma$ values of each age bin.}\label{fig:pltgb}
\end{figure}

We find that the majority of YSOs with ages $\leq$1~Myr are located coincident (in projection) with the dense parts of the filamentary molecular structures. On the other hand, older ($>$~1~Myr) YSOs are spread over the entire field. More quantitatively, we determine that 62~\% of the younger population ($\leq$1~Myr) is located inside the lowest contour level (30~K~km/s) of the $^{13}$CO(1-0) emission, while only 35~\% of the relatively older population ($>$~1~Myr) is found inside that contour level. This confirms the expectation that star formation occurs in the densest cloud regions, with the YSOs remaining associated with their natal dense gas for $\sim1\:$Myr, followed by relative motion between stars and gas or gas dispersal after this (see also \cite{2016ApJ...818...59D}). 

\begin{figure*}
 \begin{center}
\hspace{-0.1in} \includegraphics[width=6.5in]{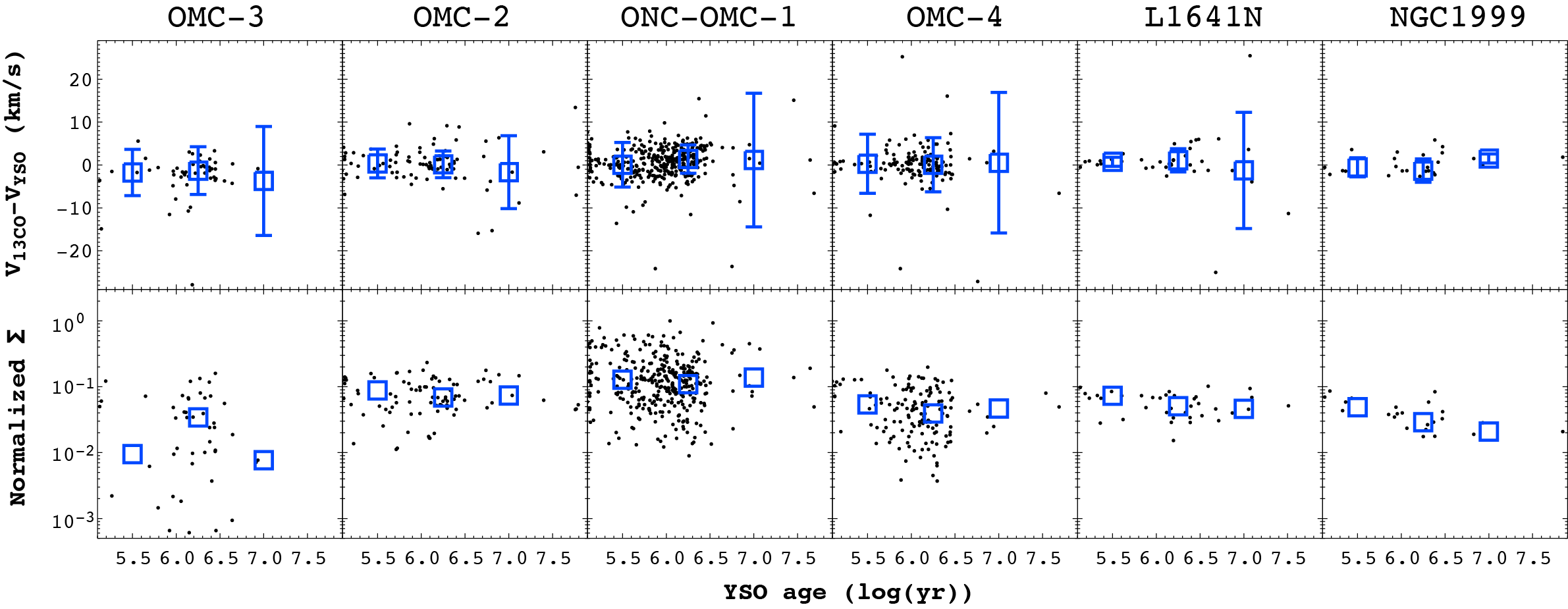}
 \end{center}
 \caption{{\bf Top:} $\Delta v$ vs.~YSO age of each region. Black dots are the YSO sources inside inner circles of figure~\ref{fig:pltgb} b). The blue boxes show the median $\Delta v$ of YSOs in three different age bins (age$\leq10^6$, $10^6 > {\rm age} \leq 10^{6.5}$ and age$>10^6.5$~years). {\bf Bottom:} $\Sigma$ vs.~age of each region. The blue boxes show the median values of normalized $^\Sigma$ in same age bins.}\label{fig:subr}
\end{figure*}

Figure~\ref{fig:pltgb}(a) shows the difference in velocity between each YSO and its local $^{13}$CO(1-0) main component, i.e. $\Delta v$, versus YSO age. The black dots are YSOs showing significant $^{13}$CO~(1-0) emission. The red dots represent those with weak ($T_{\rm mb}<3\sigma$ limit) or no $^{13}$CO emission (outside of the mapping area). We use the environmental $^{13}$CO intensities which are averaged over the CARMA-NRO beamsize ($\sim5\arcsec$) for the weak $^{13}$CO sources. The YSOs velocities with no $^{13}$CO detection have been compared to the $^{13}$CO velocity structures of the main Orion~A filament so that we average over 50~pixels { (100$\arcsec$)} in the Dec. direction as well as all pixels in the R.A. direction. The small boxes in the top left panel of figure~\ref{fig:pltgb} indicate the median values of the $\Delta v$ and the errorbars show the scatters of the data points in each bin. We choose the bin size to be 0.5~dex along the logarithmic YSO ages except for the first bin which includes all YSOs under 1~Myr.

\begin{table*}
\tbl{Observational Parameters of Stellar Clusters in Orion~A.}{%
\begin{tabular}{lccccccccc}  
\hline\noalign{\vskip3pt} 
\multicolumn{1}{c}{Source} & R.A (hh mm ss.s) & Dec. (dd mm ss.s) & r (pc) & $N_{\rm YSO}$ & $\widetilde{\Sigma}$ (g/cm$^2$)\\  
\hline\noalign{\vskip3pt} 
ONC--OMC-1 & 5 35 16.6 & -5 22 44.4 &  0.85 &  351 &  0.051 \\
     OMC-2 & 5 35 22.4 & -5 07 46.0 &  0.55 &  89  &  0.038 \\
     OMC-3 & 5 35 07.6 & -4 55 36.0 &  0.55 &  51  &  0.013 \\
     OMC-4 & 5 35 04.8 & -5 35 15.0 &  0.55 &  145 &  0.027 \\
   L1641N & 5 36 18.8 & -6 22 10.2 &  0.55 &  38  &  0.046 \\
  NGC1999 & 5 36 17.0 & -6 44 20.9 &  0.55 &  25  &  0.025 \\ 
\hline\noalign{\vskip3pt} 
\end{tabular}}\label{tb:sc}
\begin{tabnote}
\end{tabnote}
\end{table*}

Figure~\ref{fig:pltgb}(b) shows the `$\Sigma$ vs.~age' of all YSOs where the mass surface density, $\Sigma$, of each source has been derived from the major $^{13}$CO component that we use for the '$\Delta v$ vs.~age' relation. Since we cannot determine the true environmental $\Sigma$ for the YSOs without $^{13}$CO detection, we do not include them (i.e. red dotted sources in figure~\ref{fig:pltgb}(a)) for this analysis. { We show the YSO ages of non-detected $^{13}$CO(1-0) as red upward arrows (figure~\ref{fig:pltgb}(b)).}

In figure~\ref{fig:pltgb}(a) and table~\ref{tb:prmgl}, we present the '$\Delta v$ vs.~age' of YSOs that inspects the $\Delta v$ of the YSOs of age bins ending at 10$^{6.0}$, 10$^{6.5}$, 10$^{7.0}$, 10$^{7.5}$ and 10$^{8.0}$~years. We first focus on the YSOs within the CARMA-NRO CO mapped area (black colored points in figure~\ref{fig:pltgb}(a), 1,371 sources, 'Detected' in table~\ref{tb:prmgl}). These sources show a trend that older YSOs exhibit generally larger scatter in $\Delta v$ and the median values (black boxes) become more negative in older population { (see table~\ref{tb:prmgl}). The average of all YSOs (blue colored points, 1,529 sources, 'Total' in table~\ref{tb:prmgl}) present a similar trend to the black points mainly due to their large number which indicates that the IN-SYNC YSOs are mostly associated with the dense molecular structures. The YSOs outside CO maps (red colored points, 158 sources, 'Non detection' in table~\ref{tb:prmgl}) show a different distribution with $\Delta v$ values as shown in figure~\ref{fig:pltgb}(a) and table~\ref{tb:prmgl}. The red colored points of figure~\ref{fig:pltgb}(a) showing the median $\Delta v$ values (red boxes) in the age bins become more positive and the scatter gets generally bigger in the older population.}

The `$\Sigma$ vs.~age' of the entire Orion~A is shown in the bottom panel of figure~\ref{fig:pltgb} which shows that the younger YSOs are more closely associated with denser gas structures. We normalize the $\Sigma$ of each YSO position so that the maximum normalized $\Sigma$ is 1. We find that $\sim$30~\% of the YSO population with ages$\leq$1~Myr have their normalized $\Sigma$ above 0.1 while $\sim$12~\% of the YSOs with ages$>$1~Myr are above 0.1. The median values of the normalized $\Sigma$ of the age bins are shown in both figure~\ref{fig:pltgb}(b) and table~\ref{tb:prmgl}. The relatively flat trend for the bins with age$\gtrsim10^7$ years is possibly due to the small sample size while the normalized $\Sigma$ decreases along YSO evolution.

\subsubsection{'$\Delta v$ vs. age' and '$\Sigma$ vs. age' of Stellar Clusters}

In figure~\ref{fig:subr}, we inspect the local trends of six 
sub-regions corresponding to star clusters (see \cite{2018ApJS..236...25K}) marked with white circles in figure~\ref{fig:ovly}~(b). We use the same approach as that performed for the entire Orion~A region but considering larger age bin sizes so that we have three bins (age$\leq10^6$, $10^6 > {\rm age} \leq 10^{6.5}$ and age$>10^{6.5}$~years). The reason for the reduced number of bins is mainly the small number of YSOs inside the white circles (see table~\ref{tb:sc} for the number of YSOs in each circle). A typical radius (r$\sim$0.55~pc) is selected from the size of L1641N in \citet{2012ApJ...746...25N}. For convenience, we then apply the same value to all other clusters, except for the ONC-OMC-1 (r$\sim$0.85~pc) which is $\sim$1.5 times larger than all other regions. The center positions of the clusters are defined by optimal extraction \citep{1998MNRAS.296..339N} from the $^{13}$CO 0th moment map produced by \citet{2018ApJS..236...25K}. The basic properties of each young stellar cluster are summarized in table~\ref{tb:sc}.

Figure~\ref{fig:subr} and table~\ref{tb:prmlc} show the $\Delta v$ vs.~age and $\Sigma$ vs.~age of the six clusters. The $\Delta v$ vs.~age relation of ONC-OMC-1, OMC-2, OMC-4 and L1641N are qualitatively consistent with the trend found for the entire cloud so that median $\Delta v$ values are close to 0~km/s and the scatters increases for the older population. However, OMC-3 and NGC1999 show different trends to the entire Orion~A. The scatters get smaller in the older population which is possibly due to the small sample size for the old YSO population. 
The $\Sigma$ vs.~age show that the normalized $\Sigma$ of ONC--OMC-1, OMC-2, OMC-4, L1641N and NGC1999 likely decreases while increasing age, similar to the global trend. The dynamic range of the normalized $\Sigma$ does not significantly change in OMC-2, L1641N and NGC1999, i.e. different from the global trend, while ONC--OMC-1 and OMC-4 resemble the global trend (having larger dynamic range of $\Sigma$ in older population). OMC-1 shows consistently a large scatter of the normalized $\Sigma$ along the YSO ages.

\begin{table}
\tbl{Derived Parameters of Gas-YSO Association for Sub-regions}{
\begin{tabular}{l|cccccccc}
\hline\noalign{\vskip3pt} 
                        & Age bin (yrs)& $<10^6$ & $10^6$ to $10^{6.5}$ & $>10^{6.5}$ \\
\hline\noalign{\vskip3pt}
ONC-OMC-1& $\Delta v$ (km/s)         & 0.07 & 1.39 & 1.17\\
         & $\sigma (\Delta v$) (km/s)& 5.20 & 3.30 & 15.58\\
         & Normalized $\Sigma$       & 0.13 & 0.11 & 0.14\\
\hline\noalign{\vskip3pt}
OMC-2    & $\Delta v$ (km/s)         & 0.37 & 0.16 & -1.66\\
         & $\sigma (\Delta v$) (km/s)& 3.35 & 3.11 & 8.51\\
         & Normalized $\Sigma$       & 0.09 & 0.07 & 0.07\\
\hline\noalign{\vskip3pt}
OMC-3    & $\Delta v$ (km/s)         & -1.73 & -1.30 & -3.70\\
         & $\sigma (\Delta v$) (km/s)& 5.40 & 5.57 & 12.70\\
         & Normalized $\Sigma$       & 0.01 & 0.03 & 0.01 \\
\hline\noalign{\vskip3pt}
OMC-4    & $\Delta v$ (km/s)         & 0.29 & 0.06 & 0.55\\
         & $\sigma (\Delta v$) (km/s)& 6.89 & 6.32 & 16.38\\
         & Normalized $\Sigma$       & 0.05 & 0.04 & 0.05\\
\hline\noalign{\vskip3pt}
L1641N   & $\Delta v$ (km/s)         & 0.77 & 1.10 & -1.25\\
         & $\sigma (\Delta v$) (km/s)& 1.06 & 2.71 & 13.54\\
         & Normalized $\Sigma$       & 0.07 & 0.05 & 0.05\\
\hline\noalign{\vskip3pt}
NGC1999  & $\Delta v$ (km/s)         & -0.54 & -1.31 & 1.51\\
         & $\sigma (\Delta v$) (km/s)& 2.22 & 2.72 & 1.06\\
         & Normalized $\Sigma$       & 0.05 & 0.03 & 0.02\\
\hline\noalign{\vskip3pt}
\end{tabular}}\label{tb:prmlc}
\begin{tabnote}
\end{tabnote}
\end{table}

\begin{figure*}
 \begin{center}
\includegraphics[width=6.in]{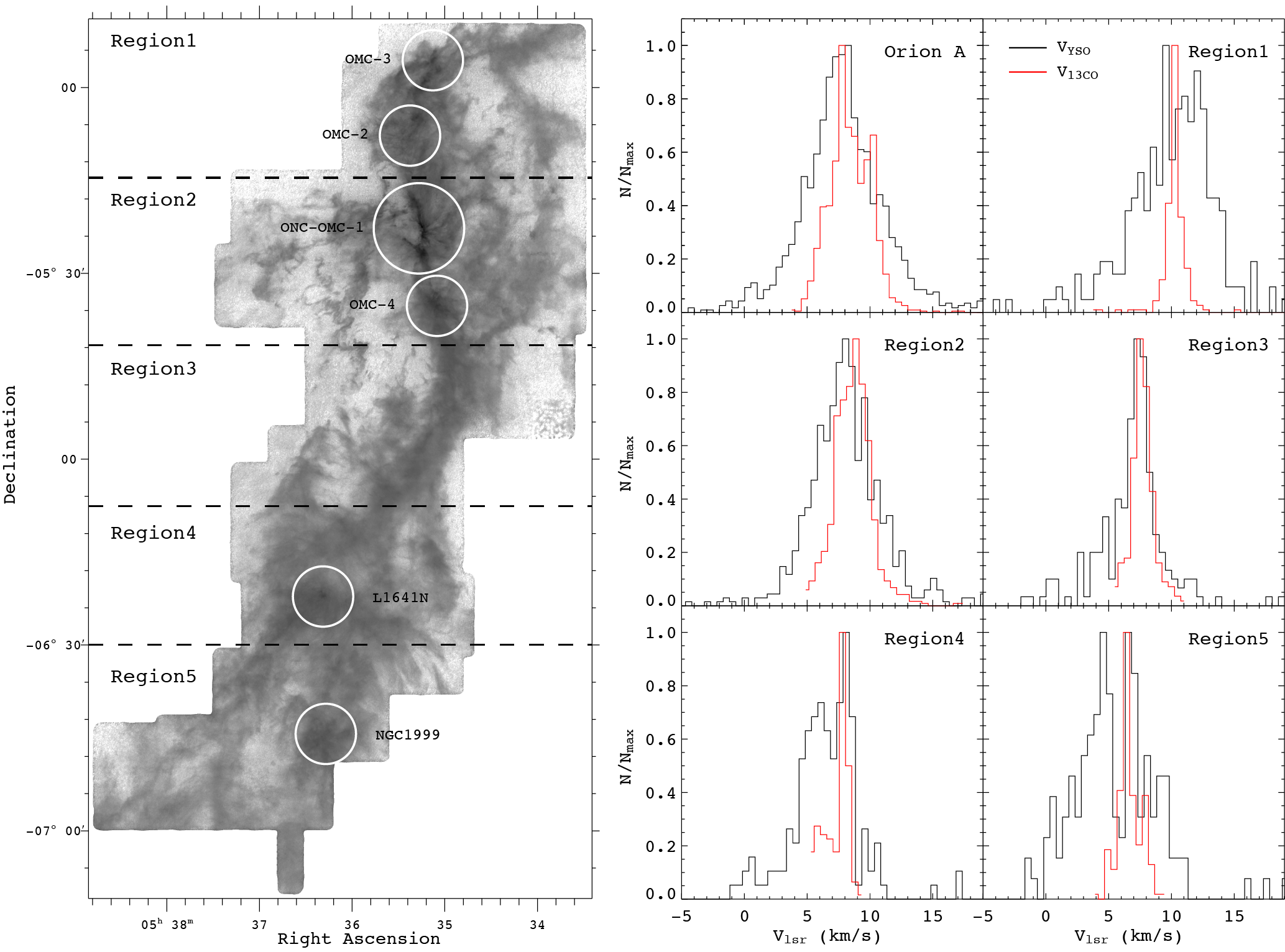}
 \end{center}
 \caption{{\bf Left:} Same as figure~\ref{fig:ovly} (right panel) but overlaying the sub-regions for $v_{\rm YSO}$ vs. $v_{\rm 13CO}$ comparison. {\bf Right: } The histograms of comparing $v_{\rm YSO}$ vs. $v_{\rm 13CO}$ of individual pixels possessing YSO locations. The inspected regions (as displayed on the top-right of each panel) are corresponding to the sub-regions of the $^{13}$CO 0th moment map while the panel named Orion~A presents the histograms for entire Orion~A area. The black and red solid lines show $v_{\rm YSO}$ and the $v_{\rm 13CO}$, respectively. We find the width of $v_{\rm 13CO}$--PDF is wider than $v_{\rm YSO}$--PDF in the norther part of Orion~A while the southern part shows similar width between both PDFs, i.e. Region~3 for the entire $v_{\rm lsr}$ range and Region~4/Region~5 for the peaks around $v_{\rm lsr}\sim$8/6~km/s. 
 This may indicate the variation of global kinematics along the Orion~A filamentary structure.}\label{fig:hist}
\end{figure*}

\subsubsection{Probability Distribution Functions of the velocities}

The Gas-YSO association along the Orion~A filament is investigated via the probability distribution function of $v_{\rm YSO}$ and $v_{\rm 13CO}$ ($v$--PDF). For this analysis, we first divide the Orion~A cloud in five regions along declination direction that include defined star clusters except Region~3 that is located between OMC-4 and L1641N. In figure~\ref{fig:hist}, we present the $^{13}$CO(1-0) 0th moment map showing the five regions on the left and the $v$--PDFs of entire Orion~A and the five regions on the right. We focus on the positions of the peaks and the Full Width Half Maximum (FWHM) of the distribution functions by performing simple multi-Gaussian fits since sometimes the peak and dispersions do not appear clearly in multi-peaked regions. We investigate the overall trend of the $v$--PDFs so that the parameters from the Gaussian fits provide us corresponding information.
The position of the peaks and the widths of the $v$--PDF provide the representative velocity structures in the inspected regions and the kinematic dispersion of the them, respectively. 

\begin{figure*}
  \begin{center}
  \includegraphics[width=6.5in]{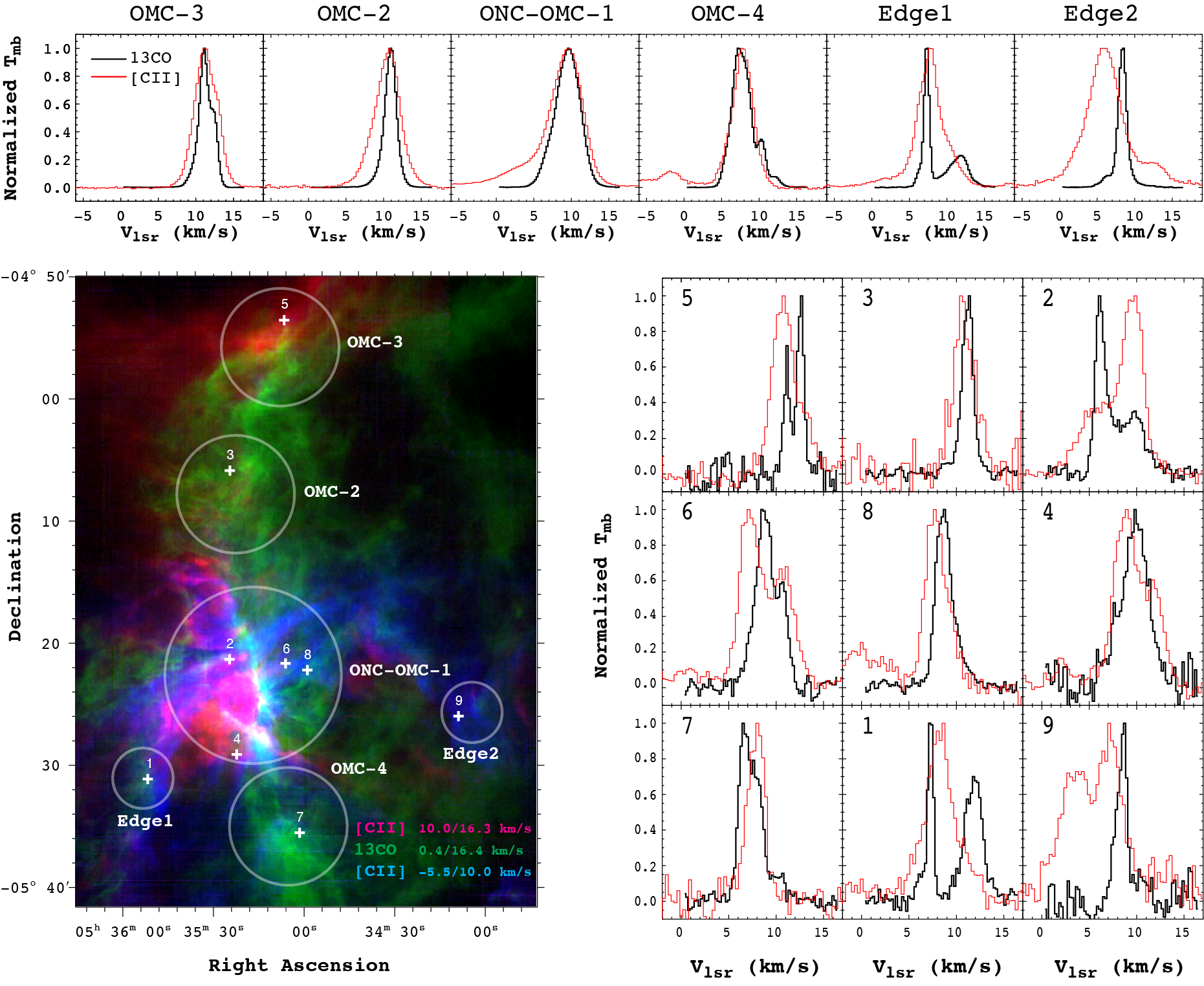}
 \end{center}
 \caption{{\bf Map:} A false-color image showing the 0th moment maps of [C{\sc ii}] (-5.5$\lesssim$v$\lesssim$10.0~km/s, blue), $^{13}$CO (green) and [C{\sc ii}] (10.0$\lesssim$v$\lesssim$16.3~km/s, red). White circles show star forming regions defined by \citet{2018ApJS..236...25K}. Edge~1 and Edge~2 regions are test regions to compare $^{13}$CO vs. [C{\sc ii}] velocity profiles at the edge of the expanding shell. The numbered crosses indicate randomly selected pixel positions for the $^{13}$CO vs. [C{\sc ii}] velocity comparison. {\bf Upper panels:} The integrated $^{13}$CO (black solid line) and [C{\sc ii}] (red solid line) velocity profiles inside white circled areas shown in the map. {\bf Right panels:} The same velocity profiles as in the upper panels but for each randomly selected pixel position marked in the map.}\label{fig:vall}
\end{figure*}

The $v$--PDFs of the entire Orion~A show single distributions for both $v_{\rm YSO}$ and $v_{\rm 13CO}$ where the velocity peaks are separated only by $\sim$0.4~km/s. 
The $v_{\rm YSO}$ distribution is wider than the $v_{\rm 13CO}$ distribution while the population of blue-shifted YSOs (compared to $v_{\rm 13CO}$) is slightly larger than for the red-shifted YSOs. The $v$--PDF peaks/FWHMs of $v_{\rm YSO}$ and $v_{\rm 13CO}$ are $\sim$8.3/7.5~km/s and $\sim$8.7/4.4~km/s, respectively. There is a second $v_{\rm 13CO}$ peak at $\sim$10.3~km/s of the global $v$--PDF which is dominated by the major $v_{\rm 13CO}$ peak of Region~1 that contains OMC-2 and OMC-3 clusters. Region~1 shows a narrow single Gaussian distribution of $v_{\rm 13CO}$ where the peak/FWHM are $\sim$10.7/0.9~km/s while the $v_{\rm YSO}$ shows two relatively wide normal distributions peaked at $\sim$10.3 (normalized probability$\sim$0.34) and $\sim$12.3~km/s (normalized probability$\sim$0.58) with FWHMs of $\sim$7.3 and 3.1~km/s, respectively. Region~2 presents mostly blue-shifted single Gaussian $v_{\rm YSO}$ distribution. The peak/FWHM of the $v_{\rm 13CO}$ and $v_{\rm YSO}$ distributions are $\sim$9.2/2.9~km/s and $\sim$8.4/5.8~km/s, respectively. Region 3, with no defined active star forming clumps/clusters, is supposed to be the most quiescent region in Orion~A. It shows the most narrow distributions for both $v_{\rm 13CO}$ and $v_{\rm YSO}$. The peak/FWHM values of $v_{\rm 13CO}$ are $\sim$8.2/0.8~km/s and $v_{\rm YSO}$ shows the values as $\sim$7.9/1.4~km/s. The $v_{\rm YSO}$ and $v_{\rm 13CO}$ of Region~4 present two distinguished populations. The peak/FWHM values of the $v_{\rm YSO}$ are $\sim$6.4/3.3~km/s and $\sim$8.6/1.0~km/s while $v_{\rm 13CO}$ show the peaks at $\sim$6.7 and 8.4~km/s for the corresponding FWHMs as $\sim$1.6 and 0.7~km/s, respectively. The $v_{\rm YSO}$ and $v_{\rm 13CO}$ peaks at $\sim$6.4 and 6.7~km/s are exclusively from L1641N cluster where the ages of most YSOs inside are smaller than 1~Myr. The Gaussian fits to the $v$--PDFs of Region~5 show three peaks of $v_{\rm YSO}$ distribution at $\sim$4.9, 6.7 and 9.6~km/s with the FWHMs of 5.8, 0.2 and 2.6~km/s, respectively. The $v_{\rm 13CO}$ distribution shows only one narrow peak at $\sim$6.9~km/s with the FWHM as $\sim$0.8~km/s. The normalized $v_{\rm 13CO}$ distribution peak of Region~5 is consistent with the $\sim$6.7~km/s peak of $v_{\rm YSO}$ distribution.

\subsection{Comparing $^{13}$CO(1-0) to [C{\sc ii}]~158~$\micron$} \label{sec:result2}

\citet{2018MNRAS.478L..54B} detected velocity and position offsets between [C{\sc ii}]~$158\,\mu$m and $^{13}$CO~(1-0) line emission maps in IRDC G035.39-00.33. From a comparison with results of numerical models, they claimed that a cloud-cloud collision could explain these offsets, and thus may be a likely formation mechanism for the IRDC. We adopt their methods to test the cloud-cloud collision scenario of star cluster formation in Orion~A using the high angular resolution [C{\sc ii}]~$158\,\mu$m (SOFIA-upGREAT) and $^{13}$CO~(1-0) (CARMA-NRO) data.

The white circles in figure~\ref{fig:vall} show the four northern clusters (ONC-OMC-1, OMC-2, OMC-3 and OMC-4) where both the [C{\sc ii}] and $^{13}$CO observations exist as well as two test regions (Edge1 and Edge2). 
\citet{2019Natur.565..618P} inspect the [C{\sc ii}] PVD of the northern part of Orion~A and compare it to a simple model of a spherically expanding shell with velocity of 13~km/s. According to their results, the locations of Edge1 and Edge2 can be assumed to be at the border of the expanding shell where the foreground velocity components are either absent or merged into the major peaks. The ONC-OMC-1 region is also consistent with the location of the border of expansion while OMC-4 is situated at the middle of the shell along the line of sight direction. The OMC-2 and OMC-3 are outside of the expanding shell thus we can expect the gas kinematics on these regions undergo minimal effects of the enormous stellar feedback by $\theta^1$~Orionis~C \citep{2019Natur.565..618P}. The $^{13}$CO(1-0) and [C{\sc ii}] emission from the six sub-regions can, thus, present the gas kinematics of cold and dense filamentary structures ($^{13}$CO) or warm envelopes ([C{\sc ii}]) in different environments with and without effects of strong stellar feedback.
 
\begin{figure}
  \begin{center}
  \includegraphics[width=3.3in]{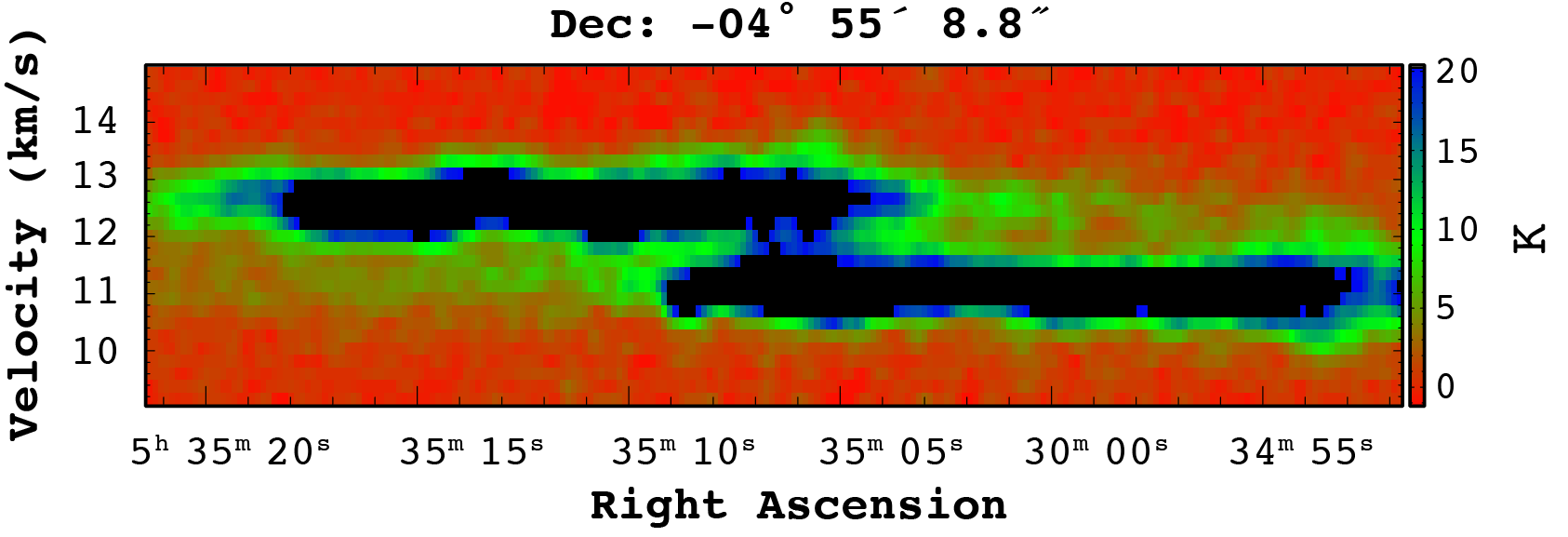}
 \end{center}
 \caption{The $^{13}$CO(1-0) PVD of a slice along the R.A. direction that includes pixel position 5 of figure~\ref{fig:vall}. This PVD shows a clear bridging effect at the position which may indicate a cloud-cloud collision in OMC-3.}\label{fig:pvd}
\end{figure}
 
The map of figure~\ref{fig:vall} shows the false-color 0th moment maps of the blue-shifted [C{\sc ii}] (blue), $^{13}$CO (green) and red-shifted [C{\sc ii}] (red) emissions. Only the northern part of the $^{13}$CO map is covered by {the} [C{\sc ii}] SOFIA-upGREAT observation. For the overlap area, we carry out a pixel by pixel velocity comparison between the $^{13}$CO and [C{\sc ii}] lines to search for observational signatures of a cloud-cloud collision. The angular resolution of [C{\sc ii}] is about three times coarser ($\sim18\arcsec$) than that of the $^{13}$CO observations ($\sim8\arcsec\times6\arcsec$). Thus, we first re-grid the $^{13}$CO map to the grid and resolution of the [C{\sc ii}] data before we perform the velocity comparison. The integrated velocity profiles of $^{13}$CO and [C{\sc ii}] inside the inspected regions (white circles in figure~\ref{fig:vall}) are presented on the upper panels of figure~\ref{fig:vall}. One also needs to note that the wider [C{\sc ii}] velocity profile than $^{13}$CO in all regions only except OMC-4 is possibly due to photoevaporation flows at the surfaces of the denser gas as we know that Orion A is a PDR to a large degree. The nine panels in the right of figure~\ref{fig:vall} show the $^{13}$CO and [C{\sc ii}] velocity profiles of randomly selected pixel positions (white crosses of figure~\ref{fig:vall}).

The top panels of Fig~\ref{fig:vall} show the integrated line emission ($^{13}$CO(1-0) -- black solid line; [C{\sc ii}] -- red solid line) inside the six sub-regions. The northernmost cluster, OMC-3, shows two velocity peaks at 11.1/12.6~km/s for $^{13}$CO and 11.2/13.2~km/s for [C{\sc ii}] based on the multi-Gaussian fit. The PVD of the CARMA-NRO $^{13}$CO data shows two clear velocity components across OMC-3 region at $\sim$11.2 and 12.6~km/s (see figure~\ref{fig:pvd}). The two velocity components are linked with a prominent ``bridge'' around (R.A., Dec.) = (5:35:10.0, -04:55:8.8) that is located inside OMC-3. The $^{13}$CO and [C{\sc ii}] emission in OMC-2 shows peak velocities of 10.9 and 10.5~km/s, respectively. Although both lines have single Gaussian shapes, the FWHM of [C{\sc ii}] emission is about twice that of the $^{13}$CO emission (4.0 vs. 1.9~km/s). The ONC-OMC-1 cluster shows wide $^{13}$CO emission, peaked at 9.7~km/s with a FWHM of 3.5~km/s. The [C{\sc ii}] emission in ONC-OMC-1 contains two merged velocity components, with peak velocities of $v_{\rm peak}\sim$5.4 and 9.6~km/s, and line widths of FWHM$\sim$10.0 and 4.4~km/s, respectively. Note that the stronger component is similar to the $^{13}$CO emission while the blue-shifted weak component is much broader. The weak [C{\sc ii}] velocity component is assumed to be a result of averaged foreground features of the expanding shell. The OMC-4 region shows clear [C{\sc ii}] foreground emission associated with the expanding shell at -1.8~km/s. The $^{13}$CO and [C{\sc ii}] major velocity components are separated only by $\sim$0.2~km/s ($v_{\rm peak,CO}\sim$7.6~km/s and $v_{\rm peak,[CII]}\sim$7.8~km/s) while the $^{13}$CO emission presents additional background features at 10.4 and 12.0~km/s. The test regions (Edge1 and Edge2) show a narrow major $^{13}$CO velocity features with broad [C{\sc ii}] multi-components.

One needs to note that \citet{2018MNRAS.478L..54B} predicted position offsets between $^{13}$CO and [C{\sc ii}] emission which are not distinguishable in the integrated line images. The detailed $^{13}$CO and [C{\sc ii}] structures enable us to investigate through a pixel by pixel comparison of both line emissions. In the bottom-right panels of Fig~\ref{fig:vall}, we show the $^{13}$CO and [C{\sc ii}] emissions at nine randomly selected positions that are inside the six sub-regions. This analysis provides an apprehensive idea of the detailed velocity structures and their variations. An extensive pixel by pixel comparison with statistical investigation of the $^{13}$CO and [C{\sc ii}] structures of Orion~A will be reported in a future paper.

In Fig~\ref{fig:vall}, the numbers move from left to right so that the numbers are based on decreasing Right Ascension. The velocity profile differences between integrated and individual pixel spectra inside each cluster show how averaging spectra smear out significant features. For example, the position of velocity peaks in Panel~2 do not match the integrated velocity of the ONC-OMC-1 region. However, all other regions show signs that the global velocity profiles (from clusters) present major and/or minor peaks agreeing with the velocity peaks of individual pixels. The spectral features of panel~1 present similar trends with the integrated emissions of the entire Edge1 area. 
A narrow $^{13}$CO peak is separated only by $\sim$2~km/s from a much wider [C{\sc ii}] peak while there is a background $^{13}$CO emission that does not seem to be connected to the narrow $^{13}$CO major peak and has a broader linewidth. Panels~2 (in ONC-OMC-1), 5 (in OMC-3) and 7 (in OMC-4) show qualitatively congruent trends that show two connected velocity components in each $^{13}$CO and [C{\sc ii}] observation where the peak velocities of the two lines agree to within $\sim$1~km/s. 
The most evident case is panel~2. The strongest peaks of the $^{13}$CO and [C{\sc ii}] emission are found at velocities of $\sim$6.0 and 9.7, respectively, while the secondary peaks are at $v_{\rm peak,13CO}\sim$11.0~km/s and $v_{\rm peak,[CII]}\sim$6.1~km/s. Panel~3 shows single Gaussian shapes of $^{13}$CO ($\sim$9--13~km/s) and [C{\sc ii}] ($\sim$8--14~km/s). Panels~4 and 6 also show two velocity components in both lines but the main peak of $^{13}$CO emission is close to the main [C{\sc ii}] velocity component. The two peaks that are separated by $\sim$1~km/s for panel~4 and $\sim$1.5~km/s for panel~6. The secondary peaks of [C{\sc ii}] line are equivalent to the secondary $^{13}$CO peaks. The full velocity ranges of [C{\sc ii}] for both panels~4 and 6 include the $^{13}$CO velocity components. Panel 8 shows single Gaussian shapes of $^{13}$CO and [C{\sc ii}] emission but the velocity range of $^{13}$CO emission does not completely overlap with the [C{\sc ii}] emission. The pixel numbered~9 is situated at the edge of the expanding shell (inside Edge2 region). The $^{13}$CO emission features of the entire Edge2 region and the pixel numbered~9 quantitatively agree with each other for the peak velocity and the narrow velocity ranges. This may indicate that dense molecular structure in/around the Edge2 area is kinematically quiescent. The [C{\sc ii}] emission in panel~9 shows a significant secondary feature that is possibly associated with the expanding shell moving toward us, while the integrated [C{\sc ii}] emission inside the Edge2 region shows a secondary feature which may move away from us. This implies that the warm envelope traced by [C{\sc ii}] has extreme kinematic variations in/around the Edge2 area compared to the $^{13}$CO emission. 

\begin{figure}
 \begin{center}
\includegraphics[width=3.1in]{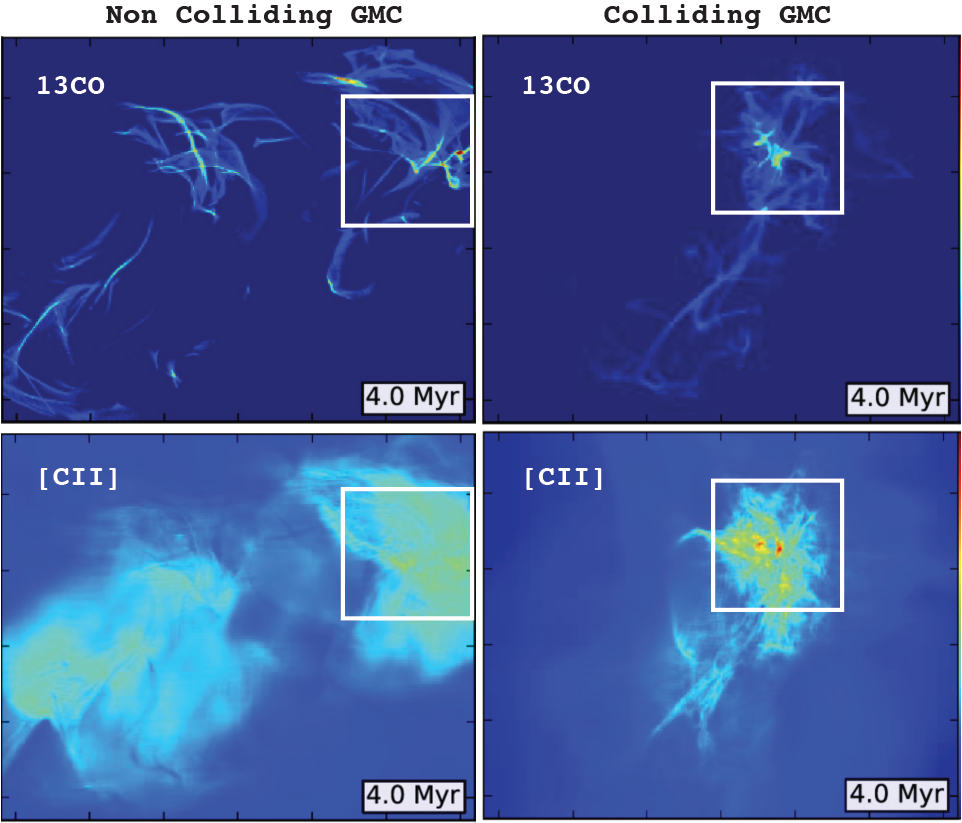}
 \end{center}
 \caption{Synthetic line emissions ($^{13}$CO - top \& [C{\sc ii}] - bottom) of non-colliding GMCs (left) and colliding GMCs (right) after 4~Myrs \citep{2017ApJ...835..137W}. The white rectangles are the investigated locations for the analyses done in figure~\ref{fig:sim1} and figure~\ref{fig:sim2}.}\label{fig:sim1b}
\end{figure}

\begin{figure}
 \begin{center}
\includegraphics[width=3.1in]{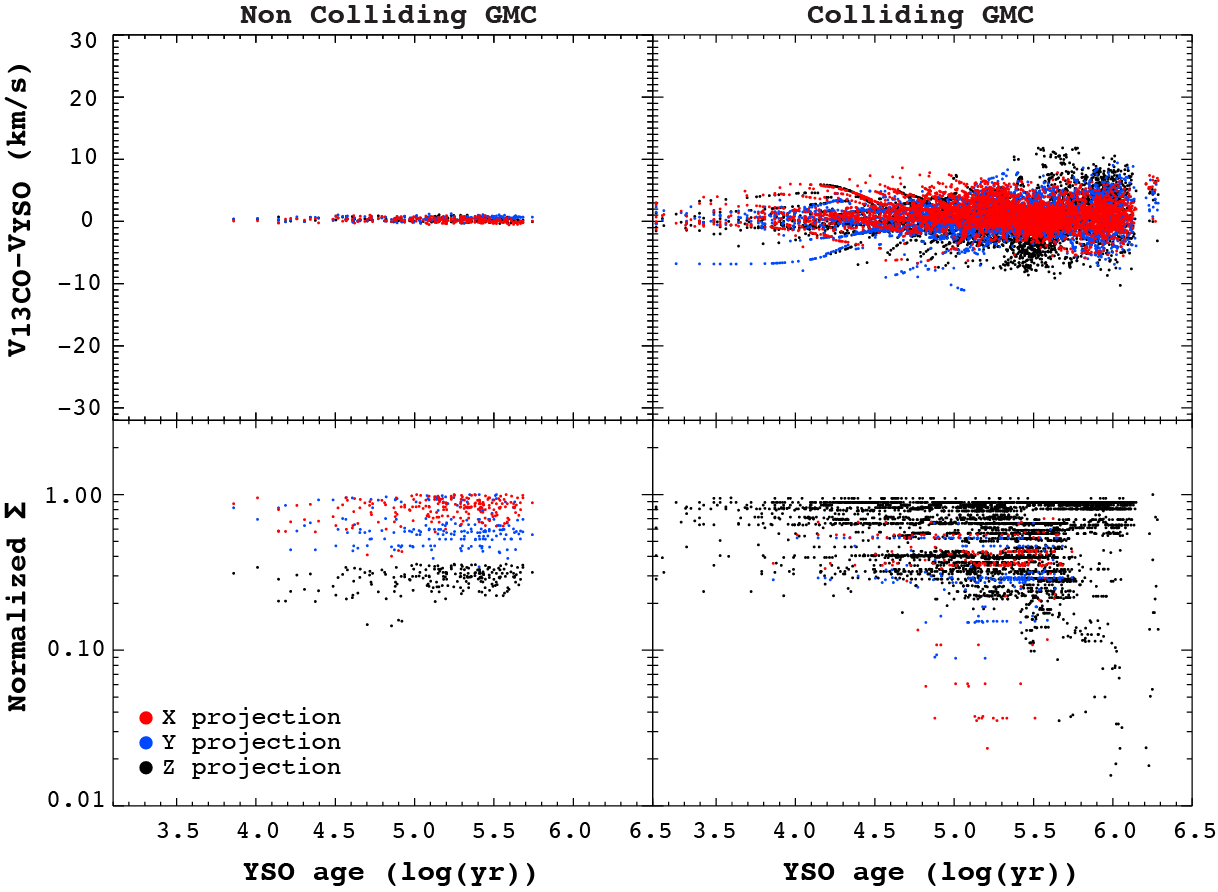}
 \end{center}
 \caption{{\bf Top:} The simulated `$\Delta v$ vs.~age' of non-colliding GMC (left) and colliding GMC (right), respectively. The results of different projects are displayed in red (X), blue (Y) and black (Z). {\bf Bottom:} Same as the top plots but for normalized $\Sigma$ vs.~age.}\label{fig:sim1}
\end{figure}

\section{Discussion}\label{sec:discussions}

In this study, we present kinematic properties of Orion~A by analyzing three independent survey data sets -- SDSS~III IN-SYNC, CARMA--NRO ORION $^{13}$CO(1-0) and SOFIA-upGREAT 158$\micron$ [C{\sc ii}] surveys. These observations trace distinct targets and their properties. The IN-SYNC data provide us with the age and radial velocities of YSOs while CARMA--NRO $^{13}$CO(1-0) data are sensitive to the gas motion of the coldest and densest parts of molecular structures (e.g. clumps and cores). The SOFIA 158~$\micron$ [C{\sc ii}] fine-structure line measures the energy input into the dense medium possessing insufficient optical depth to shield UV radiation. Here, we discuss the further indication of the analytic results and also compare them to the theoretical models.

\subsection{Trends}

We have investigated the Gas-YSO association in Orion~A via analyzing $\Delta v$ vs.~age, $\Sigma$ vs.~age, and $v$-PDFs of the entire Orion~A region as well as for 6 sub-regions (young stellar clusters). The analysis of Gas-YSO association reveals a trend where dense molecular structures and YSOs lose kinematic adherence with evolution, especially in the northern part of Orion~A.

\subsection{Comparison with modelling}

We compare the observational results of Gas-YSO associations with those measured from numerical simulations of star cluster formation. To investigate the likelihood of a cloud-cloud collision formation scenario for Orion A, we utilize magnetohydrodynamics (MHD) simulations from \authorcite{2017ApJ...835..137W} (\yearcite{2017ApJ...835..137W, 2017ApJ...841...88W}), who developed models comparing non-colliding and colliding GMCs. The \citet{2017ApJ...835..137W} simulations focused on dense molecular structures and synthetic observations. We additionally utilize the \citet{2017ApJ...841...88W} results which involve modelling of stellar dynamics. { Figure~\ref{fig:sim1b}} present the snapshots of the simulation 4~Myrs after the events (collapsing/colliding) start. 
{Interestingly, the simulated $^{13}$CO distribution roughly resembles the global shape of the Orion A molecular cloud, i.e. its filamentary shape with a denser northern part, although the simulation was not intended to reproduce the Orion A cloud. It is worth noting that the off-centered collision preferentially creates such a cloud structure.} 
{ While MHD simulations of low-mass star forming regions are capable of producing filamentary structures (e.g. \cite{2016ApJ...833...10I}), these simulations result in clouds with much lower column density than observed in the Orion A region (by about two orders of magnitude).}

{ Figure~\ref{fig:sim1}} present the ‘$\Delta v$ vs.~age’ and ‘$\Sigma$ vs.~age’ relations for the non-colliding and colliding GMCs from the simulated data. The $^{13}$CO and [C{\sc ii}] simulations presented in the current work are derived from \citet{2017ApJ...835..137W} by post-processing individual snapshots. Using the astrochemical code {\sc 3d-pdr}\footnote{https://uclchem.github.io/3dpdr.html} \citep{2012MNRAS.427.2100B} and the PDR methodology of connecting the local total H-nucleus number density with a most probable value of effective visual extinction from \citet{2017ApJ...850...23B} {(see also \cite{2015ApJ...811...56W})}, we perform synthetic observations of GMC simulations described in \citet{2017ApJ...835..137W}.
The white rectangles { in figure~\ref{fig:sim1b}} show the location where we extract the data points (their size is $\sim16\times16$~pc$^2$, note that entire CARMA--NRO Orion survey field is $\sim$16~pc; \cite{2018ApJS..236...25K}). We find that the cloud-cloud collision scenario qualitatively reproduces better the Gas-YSO association of the entire Orion~A (cf. figure~\ref{fig:pltgb}; $\widetilde{\Delta v}\sim$0~km/s, larger $\Delta v$ scatter at older population, decreasing $\widetilde{\Sigma}$ along evolution and, larger $\Sigma$ scatter at older population) when compared to the non-colliding scenario. It should be noted that the numerical codes underestimate the increasing $\Delta v$ scatter of the YSOs since the spatial resolution of the simulations are not enough to resolve the close encounters \citep{2017ApJ...841...88W}. Thus, the models provide lower limits but the qualitative results, i.e. the trends, shown here should be met.
The non-colliding GMC model shows that the stars are consistently well associated with the dense molecular clouds, thus no active dispersion happens, while the cloud-cloud collision generates large dispersion throughout its dynamical evolution. These simulated trends are consistent for different projections as seen in { figure~\ref{fig:sim1}}.  

As seen in figure \ref{fig:subr}, ONC-OMC-1 and OMC-4 clusters show the clearest trends resembling the colliding GMC cases while OMC-2 arguably favors the cloud-cloud collision scenario only from the trend of $\Delta v$ vs.~age. The trends for OMC-3 are very different from any other sub-regions showing smaller $\Delta v$ vs.~age scatter for the older population and a large scatter in $\Sigma$ vs.~age without any specific trend.
This could indicate a young dynamic age of the region where two or more star forming clouds are interacting, merging, and thereby triggering more star-formation. 
The NGC1999 region is relatively quiescent and more consistent with the non-colliding GMC case. On the other hand, L1641N shows that a trend of $\Delta v$ vs.~age that is more consistent with the cloud-cloud collision scenario while the tendency of $\Sigma$ vs.~age is unclear.

\begin{figure}
  \begin{center}
  \includegraphics[width=3.2in]{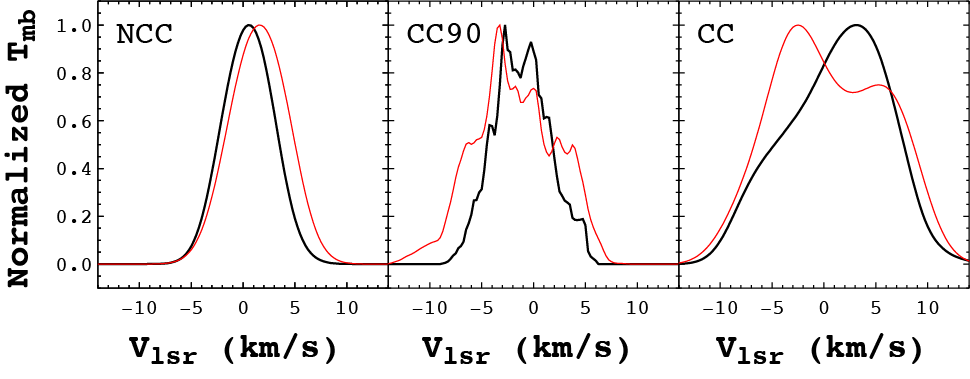}
 \end{center}
 \caption{Synthetic $^{13}$CO (black solid line) and [C{\sc ii}] (red solid line) velocity profiles based on the simulations shown in { figure~\ref{fig:sim1b}}. The panels named 'NCC', 'CC90' and 'CC' represent the simulation results (based on \cite{2018MNRAS.478L..54B}) of non-colliding clouds, colliding clouds with edge-on view and colliding clouds with face-on view, respectively. These profiles are extracted from a thin slice for each case representing the local conditions at the densest region of the cloud (see $\S$\ref{sec:discussions} for detail).} \label{fig:sim2}
\end{figure}

The possibility of cloud-cloud collision in the northern part of Orion~A is collated by comparing the observed line profiles of the $^{13}$CO and [C{\sc ii}] emission to theoretical models based on \citet{2018MNRAS.478L..54B}. In particular, we focus on sub-regions within non-colliding and colliding clouds at dynamical times of $t=4\,{\rm Myr}$ (see also \cite{2018MNRAS.478L..54B}) containing {dense filaments} and we perform radiative transfer calculations for the lines of [C{\sc ii}] $158\,\mu$m and {$^{13}$CO~(1-0)} (covering the white rectangles in figure~\ref{fig:sim1}). To obtain the level populations of $^{13}$CO, we assume an isotopologue abundance ratio of $^{12}$CO/$^{13}$CO$=60$, close to the value of 69.6 used in Sect. 2.2. 
It can be seen that there is a velocity offset between the peaks of these two emission lines, indicating a cloud-cloud collision scenario in agreement with \citet{2018MNRAS.478L..54B} who performed detailed, full 3D simulations in a similar case. 

The panels entitled `NCC', `CC90' and `CC' in figure~\ref{fig:sim2} show the [C{\sc ii}] and $^{13}$CO spectra of non-colliding clouds, colliding clouds {from an} edge-on view, and colliding clouds {from a} face-on view, respectively. NCC shows single Gaussian profiles for both lines, with small velocity offsets which can be associated with turbulent motions in a self-gravitationally collapsing cloud. The CC model shows a clear offset between $^{13}$CO and [C{\sc ii}] main velocity peaks while the linewidth of [C{\sc ii}] embraces $^{13}$CO in both CC and CC90 cases. The densest part of merging clouds can be traced at the major $^{13}$CO peak while the [C{\sc ii}] lines may trace the envelopes around the merging interface. CC90 shows complicated features which can naturally arise from the turbulent motion of the gas. We summarize the observational signals of cloud-cloud collision by the $^{13}$CO vs.~[C{\sc ii}] comparison as follows. 
The trends between position and velocity offsets observed in Orion~A (see figure~\ref{fig:vall}) are similar to those suggested in \citet{2018MNRAS.478L..54B}. In addition, a [C{\sc ii}] linewidth larger than the $^{13}$CO one where the [C{\sc ii}] covers entire $^{13}$CO line (we refer to as [C{\sc ii}] embracing) suggest a cloud-cloud collision scenario.  

The spectra from the theoretical models were extracted from a 2~pc width strip across the densest part of the cloud for each colliding scenario as well as along the line of sight \citep{2018MNRAS.478L..54B}. The entire simulated cloud structure has its effective radius, $r_{\rm eff}\gtrsim20$~pc so the strip is considered as a thin slice of the cloud. Since the 'thin slice' represents the local environment of each scenario, the proper observational analogue would be to analyze individual spectral pixels rather than the averaged spectra of an entire cluster.
This aspect is presented in the upper panel of figure~\ref{fig:vall}. The integrated $^{13}$CO and [C{\sc ii}] line profiles do not show clear trends of cloud-cloud collision in any of the northern sub-regions, except the larger [C{\sc ii}] linewidth, compared to the $^{13}$CO linewidth, in OMC-1--ONC, OMC-2 and, OMC-3 regions. The line profiles of individual pixel positions provide hints of non-colliding and/or colliding GMC scenarios in different regions. Panels 2, 3, 4, 5, and 6 show consistency with the CC model, while panels 7 and 8 arguably show single Gaussian profiles for both $^{13}$CO and [C{\sc ii}] with certain velocity offsets without the [C{\sc ii}] embracing which may favor of the NCC model. The velocity profiles of shell locations 1 and 9 resemble the integrated line emission of the entire test regions. Based on the analytic results, we assume that the cloud-cloud collision in the northern part of Orion~A may be an important mechanism for the ongoing star cluster formation. This is consistent with an ammonia line survey toward the cores of Orion~A cloud by \citet{2017ApJ...846..144K}. They found the most cores are not gravitationally bound but created and stabilized by external pressure.
One needs to note that such mechanism can be also triggered by expending PDR shells located near to the OMC-3 region. A well known expanding bubble, NGC~1977, is located close to the northern direction of OMC-3.
A more extended analysis of the pixel by pixel $^{13}$CO vs.~[C{\sc ii}] comparison will help to statistically confirm or refute the cloud-cloud collision scenario in the Orion~A region in a forthcoming work.

\subsection{Concluding remarks}

We have investigated the kinematic properties of various structures in the Orion~A cloud in order to examine the star cluster formation mechanisms. Even with the qualitatively consistent observational evidences of cloud-cloud collision as an important star cluster forming mechanisms in Orion~A, one needs further interpretation to 
this as the dominant process of the star cluster formation. Similar to \citet{2019ApJ...875..162F} and \citet{2019Natur.565..618P}, the kinematic contribution of stellar feedback in Orion~A is significant enough to affect the entire system where the feedback from OB association can even create the Orion~A filament \citep{1987ApJ...312L..45B}. For instance, the gas around the older YSOs can have been dispersed by external processes or feedbacks so that the loosening Gas-YSO association is generated by such mechanisms. Since the simulations do not contain stellar feedback, the theoretical models including strong stellar feedback are needed to confirm the observational evidences of cloud-cloud collision. The large structure dynamics involving multi-number of molecular structures can include, but is not limited to gravitational converging flows from a rotating sheet \citep{2007ApJ...654..988H}, and the oscillating dense filamentary structures which can eject star system from the mother clouds, so called slingshot mechanism \citep{2016A&A...590A...2S}. { These scenarios can explain the larger $\Delta v$ scatter at older YSO population by the global motions of the molecular clouds but are unable to address the velocity offset between the $^{13}$CO and [C{\sc ii}] emission.} With careful investigations involving the large scale dynamics with various levels of stellar feedback input, we will be able to have more complete understanding about what observations tell us about the Galactic star cluster formation processes. 

\begin{ack}
The authors thank an anonymous referee for constructive comments that help to improve the manuscript. The authors also thank B.-G. Andersson, H. Arce, J. Carpenter, J. Jackson, Y. Lee, T. Le Ngoc, W. Reach, A. Soam and W. Vacca for helpful discussions. WL acknowledges supports from SOFIA Science Center.
This work is based on observations made with the NASA/DLR Stratospheric Observatory for Infrared Astronomy (SOFIA). SOFIA is jointly operated by the Universities Space Research Association, Inc. (USRA), under NASA contract NAS2-97001, and the Deutsches SOFIA Institut (DSI) under DLR contract 50 OK 0901 to the University of Stuttgart.
GREAT, used for the [C{\sc ii}] observations, is a development by the MPI f\"{u}r Radioastronomie and the KOSMA / Universit\"{a}t zu K\"{o}ln, in cooperation with the MPI f\"{u}r Sonnensystemforschung and the DLR Institut f\"{u}r Planetenforschung.
Part of this research was carried out at the Jet Propulsion Laboratory, California Institute of Technology, under a contract with the National Aeronautics and Space Administration. This work was carried out as one of the large projects of the Nobeyama Radio Observatory (NRO), which is a branch of the National Astronomical Observatory of Japan, National Institute of Natural Sciences. We thank the NRO staff for both operating the 45 m and helping us with the data reduction. This work was financially supported by Grant-in-Aid for Scientific Research (Nos. 17H02863, 17H01118).
\end{ack}

\appendix
\section{Comparing Multi-component Gaussian fitters}\label{sec:app}

\begin{figure*}
  \begin{center}
  \includegraphics[width=7in]{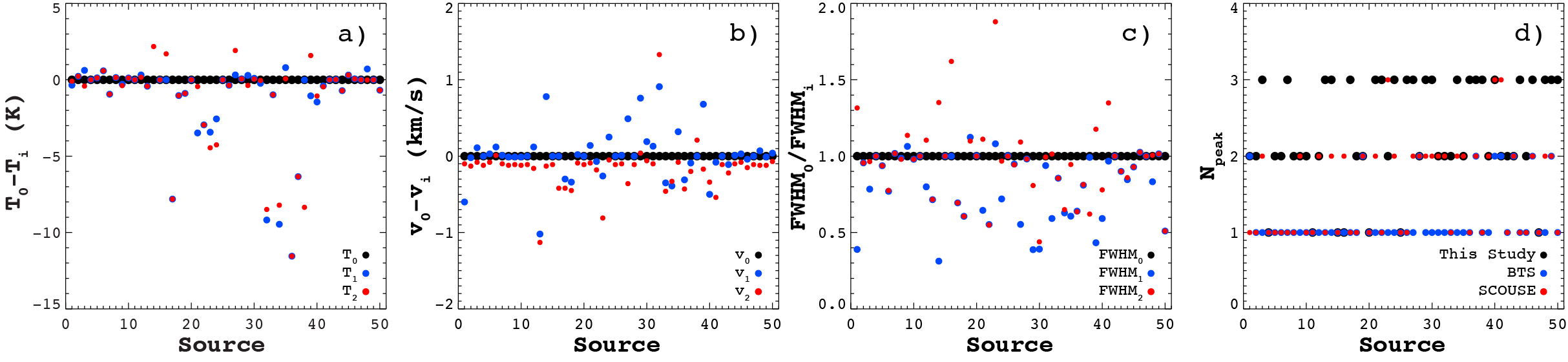}
 \end{center}
 \caption{Comparisons of the parameters derived from the three different Multi-Gaussian fitters. As for table~\ref{tb:gauss}, we focus only on the parameters of the main peaks. Panel a) shows the comparison of peak temperatures of the fitters. $T_{\rm 0}$ (black dots) indicates the peak temperature from the Gaussian fitter of this study. $T_{\rm 1}$ (blue dots) and $T_{\rm 2}$ (red dots) represent the peak temperatures of BTS and SCOUSE, respectively. Panel b) and c) are same as panel a) but for the comparisons of the central velocities and FWHMs. Panel d) shows the number of peaks from the three methods.} \label{fig:gausscomp}
\end{figure*}

In this section, we introduce the derived parameters of the three different multi-component line fitters of this study, SCOUSE, and BTS. In table~\ref{tb:gauss} and figure~\ref{fig:gausscomp}, we present the parameters of the main components only with the number of components each fitter derived. As shown in figure~\ref{fig:gausscomp}, we compare the three Gaussian parameters, i.e. the peak intensities, central velocities, and FWHMs of the main components, among the fitters while we use the parameters of this study as the references. The 42 out of 50 sample YSOs show that the peak intensities of BTS and/or SCOUSE agree within 2~K. The other 8 sources show that both BTS and SCOUSE agree each other for the peak intensities within 2~K while they differ from the intensities from this study by $\sim$3--12~K. The FWHMs also vary by about a factor of two at the most extreme case. The central velocities are agreeing quite well (all within 1~km/s) so that our main point, investigating the difference in the velocities, is very robust.

The fitter of this study tends to have more number of peaks than BTS and SCOUSE. This is mainly due to their approaches to multi-components having close central velocities. The multi-Gaussian fitter of this study assumes that such close components are results of very close gas structures while BTS and SCOUSE sometimes assume those as results of self-absorption lines. This can also explain the several of the much higher intensity peaks in BTS and SCOUSE than those of this study. One needs to note that we use $^{13}$CO(1-0) emission line which is optically thin in general where the effects of self-absorption are rare. 

\renewcommand{\arraystretch}{0.70}    
\begin{table*}
\tbl{Derived Parameters of the Multi-Gaussian Fits to Fifty of Randomly Selected Sources}{\begin{tabular}{lccccccccccccc}  
\hline\noalign{\vskip3pt} 
\multicolumn{1}{c}{Source} & \multicolumn{4}{c}{This Study} & \multicolumn{4}{c}{BTS} & \multicolumn{4}{c}{SCOUSE} \\\cmidrule(lr){2-5} \cmidrule(lr){6-9} \cmidrule(lr){10-13}
  & $N_{\rm peak}$ & $T_{\rm peak}$ & $v_{\rm peak}$ & FWHM & $N_{\rm peak}$ & $T_{\rm peak}$ & $v_{\rm peak}$ & FWHM & $N_{\rm peak}$ & $T_{\rm peak}$ & $v_{\rm peak}$ & FWHM \\
  & & (K) & (km/s) & (km/s) & & (K) & (km/s) & (km/s) & & (K) & (km/s) & (km/s) \\
\hline\noalign{\vskip3pt}
1 &	2	&12.93	&7.28	&1.00	&2	&13.28	&7.88	&2.57	&1	&13.02	&7.38	&0.76 \\
2 &	2	&16.64	&6.85	&1.53	&1	&16.41	&6.87	&1.60	&1	&16.41	&6.98	&1.60 \\
3 &	3	&18.81	&6.94	&1.09	&1	&18.19	&6.83	&1.39	&2	&19.22	&7.02	&1.13 \\
4 &	1	&17.93	&6.77	&1.34	&1	&17.93	&6.77	&1.34	&1	&17.93	&6.89	&1.34 \\
5 &	2	&17.93	&6.96	&1.35	&1	&17.80	&6.93	&1.44	&1	&17.80	&7.04	&1.44 \\
6 &	2	&18.77	&6.95	&1.07	&1	&18.18	&6.83	&1.39	&1	&18.18	&6.94	&1.38 \\
7 &	3	&6.08	&6.60	&2.81	&1	&7.02	&6.59	&2.76	&1	&7.02	&6.70	&2.76 \\
8 &	2	&16.79	&6.92	&1.46	&1	&16.64	&6.93	&1.48	&1	&16.64	&7.04	&1.49 \\
9 &	2	&5.92	&4.17	&1.00	&1	&6.21	&4.18	&0.94	&2	&6.30	&4.28	&0.88 \\
10&	2	&17.01	&6.93	&1.43	&1	&16.89	&6.94	&1.46	&1	&16.89	&7.05	&1.46 \\
11&	1	&21.09	&8.00	&1.77	&1	&21.09	&8.00	&1.77	&1	&21.09	&8.11	&1.77 \\
12&	2	&5.13	&14.21	&2.11	&1	&4.81	&14.09	&2.64	&2	&4.96	&14.37	&1.91 \\
13&	3	&4.15	&7.84	&3.84	&1	&4.57	&8.86	&5.37	&1	&4.57	&8.97	&5.36 \\
14&	3	&23.72	&10.29	&1.00	&1	&7.89	&9.51	&3.20	&2	&21.54	&10.42	&0.74 \\
15&	1	&4.07	&9.45	&4.50	&1	&4.07	&9.45	&4.50	&1	&4.07	&9.56	&4.50 \\
16&	1	&7.88	&10.41	&3.16	&1	&7.88	&10.41	&3.16	&2	&6.18	&10.83	&1.95 \\
17&	3	&27.85	&11.33	&1.00	&1	&35.66	&11.63	&1.44	&1	&35.66	&11.75	&1.44 \\
18&	2	&20.56	&7.61	&1.00	&1	&21.59	&7.95	&1.65	&1	&21.59	&8.06	&1.65 \\
19&	2	&13.49	&8.56	&1.00	&2	&14.38	&8.54	&0.89   &2	&14.38	&8.65	&0.91 \\
20&	1	&3.78	&9.28	&4.15	&1	&3.78	&9.28	&4.14	&1	&3.78	&9.39	&4.15 \\
21&	3	&19.41	&10.43	&1.00	&1	&22.89	&10.29	&1.55	&2	&19.85	&10.52	&0.90 \\
22&	3	&13.94	&9.35	&1.00	&1	&16.89	&9.42	&1.81	&1	&16.89	&9.53	&1.82 \\
23&	2	&12.01	&9.43	&2.67	&1	&15.44	&9.69	&2.47	&3	&16.46	&10.24	&1.42 \\
24&	3	&26.26	&11.44	&1.49	&1	&28.82	&11.19	&2.07	&2	&30.52	&11.49	&1.54 \\
25&	1	&11.82	&9.58	&2.88	&1	&11.82	&9.58	&2.87	&1	&11.82	&9.69	&2.88 \\
26&	3	&28.98	&10.95	&2.01	&1	&29.34	&10.94	&2.12	&1	&29.34	&11.05	&2.13 \\
27&	3	&9.45	&10.40	&1.77	&1	&9.13	&9.91	&3.20	&2	&7.53	&10.76	&1.62 \\
28&	2	&8.81	&8.52	&1.05	&2	&8.75	&8.52	&1.07	&2	&8.75	&8.63	&1.07 \\
29&	3	&4.43	&13.27	&1.47	&1	&4.14	&12.51	&3.79	&2	&4.80	&13.23	&1.82 \\
30&	3	&3.66	&10.75	&1.15	&1	&3.56	&10.56	&2.94	&2	&3.61	&10.81	&2.62 \\
31&	2	&28.21	&7.94	&3.36	&1	&28.44	&7.81	&3.58	&2	&28.25	&8.04	&3.38 \\
32&	2	&19.52	&10.38	&1.45	&1	&28.70	&9.47	&2.45	&2	&28.01	&9.05	&1.43 \\
33&	2	&22.21	&9.17	&3.08	&1	&23.19	&9.52	&3.60	&1	&23.19	&9.63	&3.60 \\
34&	3	&9.48	&8.86	&1.91	&1	&18.94	&9.25	&3.04	&2	&17.69	&9.19	&2.94 \\
35&	2	&6.22	&8.32	&1.56	&1	&5.42	&8.00	&2.57	&2	&6.15	&8.40	&1.65 \\
36&	3	&22.46	&8.95	&1.19	&1	&34.01	&9.26	&1.86	&1	&34.01	&9.38	&1.87 \\
37&	3	&15.99	&7.74	&1.62	&2	&22.33	&7.83	&2.00	&2	&22.33	&7.94	&1.99 \\
38&	3	&25.96	&7.88	&1.31	&1	&25.97	&7.88	&1.32	&1	&34.31	&7.67	&2.11 \\
39&	2	&19.06	&9.33	&1.00	&1	&20.11	&8.65	&2.31	&2	&17.47	&9.50	&0.85 \\
40&	3	&11.49	&5.68	&1.13	&2	&12.94	&6.18	&1.91	&3	&12.55	&6.02	&1.45 \\
41&	2	&25.83	&9.03	&3.25	&2	&26.23	&9.11	&3.36	&3	&26.29	&9.57	&2.41 \\
42&	2	&20.11	&9.08	&1.39	&1	&20.11	&9.08	&1.39	&1	&20.11	&9.19	&1.39 \\
43&	2	&8.28	&7.05	&2.26	&2	&8.26	&7.16	&2.51	&2	&8.26	&7.27	&2.51 \\
44&	3	&14.33	&10.68	&1.10	&1	&15.04	&10.67	&1.30	&1	&15.04	&10.78	&1.28 \\
45&	2	&14.20	&9.08	&1.18	&1	&13.89	&9.05	&1.27	&1	&13.89	&9.16	&1.27 \\
46&	3	&9.16	&8.37	&2.40	&2	&9.10	&8.41	&2.34	&2	&9.10	&8.52	&2.34 \\
47&	1	&20.27	&8.27	&1.98	&1	&20.27	&8.27	&1.98	&1	&20.27	&8.38	&1.98 \\
48&	3	&10.87	&8.40	&1.69	&1	&10.16	&8.33	&2.03	&2	&10.88	&8.52	&1.68 \\
49&	3	&9.90	&8.53	&1.98	&2	&9.90	&8.54	&1.95	&2	&9.90	&8.65	&1.95 \\
50&	3	&7.21	&8.09	&1.31	&1	&7.89	&8.05	&2.57	&1	&7.89	&8.16	&2.57 \\
\hline\noalign{\vskip3pt} 
\end{tabular}}\label{tb:gauss}
\begin{tabnote}
\end{tabnote}
\end{table*}
\renewcommand{\arraystretch}{1.0}  

\end{document}